%% file: arxivmain.tex
\documentclass[sigplan]{acmart}

\acmConference[arXiv Preprint]{Work in Progress}{Work in Progress}{Palo Alto, California}
\acmISBN{} % \acmISBN{978-x-xxxx-xxxx-x/YY/MM}
\acmDOI{} % \acmDOI{10.1145/nnnnnnn.nnnnnnn}
\startPage{1}

\setcopyright{none}
\usepackage{booktabs}   %% For formal tables:
\usepackage{subcaption} %% For complex figures with subfigures/subcaptions
\usepackage{listings}
\usepackage{booktabs}   %% For formal tables: http://ctan.org/pkg/booktabs
\usepackage{multirow}   %% For tables with merged columns/rows
\usepackage{array}
\usepackage{dcolumn}
\usepackage{hyperref}
\usepackage{dsfont}
\usepackage{multirow}
\hypersetup{
    colorlinks=true,
    linkcolor=blue,
    filecolor=magenta,
    urlcolor=cyan,
}

\newcolumntype{d}[1]{D{.}{.}{#1}}
\usepackage{graphicx}

\usepackage{amsmath}

\lstdefinelanguage{Pseudo}{
	tabsize=2,
	basewidth={0.55em, 0.4em},%
	numbers=left,
	showspaces=false,
	keywordstyle=\bfseries,
	breaklines=true,
	columns=fixed,
	basicstyle=\fontsize{9}{9}\selectfont\ttfamily,
	morestring=[b]",
	morestring=[b]',
	morecomment=[l]{//},
	morecomment=[s]{/*}{*/},
	escapeinside={@@}{@@},
	commentstyle=\itshape\color{green!40!black},
	keywordstyle=\bfseries,
	morekeywords={function, input, output, if, while, by, par, to, for, all, each, let, in, else, break, then, end, return},
	belowskip=-7pt,
	aboveskip=-7pt
}

\lstdefinelanguage{assembly}{
	tabsize=2,
	basewidth={0.55em, 0.4em},%
	numbers=none,
	showspaces=false,
	keywordstyle=\bfseries,
	breaklines=true,
	columns=fixed,
	basicstyle=\fontsize{9}{9}\selectfont\ttfamily,
	morestring=[b]",
	morestring=[b]',
	morecomment=[l]{//},
	morecomment=[s]{/*}{*/},
	escapeinside={@@}{@@},
	commentstyle=\itshape\color{green!40!black},
	keywordstyle=\bfseries,
	morekeywords={function, input, output, if, by, par, loadTile, to, for, all, each, let, in, else, break, then, end, return}
}

\lstdefinelanguage{Spatial}{
	basicstyle=\fontsize{9}{9}\selectfont\ttfamily,frame=tlbr,framesep=4pt,framerule=0pt,
	tabsize=2,
	basewidth={0.55em, 0.4em},%
	numbers=left,
	showspaces=false,  
	keywordstyle=\bfseries,   
	breaklines=true,
	columns=fixed,
	firstnumber=auto,
	showstringspaces=false,
	escapechar=@,
	escapeinside={(*@}{@*)},
	morestring=[b]",
	morestring=[b]',
	morecomment=[l]{//},
	morecomment=[s]{/*}{*/},
	backgroundcolor=\color{backcolour},   
	commentstyle=\color{codegreen}, %\bfseries,
	belowskip=-10pt,
	aboveskip=-7pt,
	numberstyle=\tiny\color{codegray},
	stringstyle=\color{codepurple},
	keywordstyle=[2]\color{blue},
	keywords=[2]{val, def},
	keywordstyle=[3]\color{blue}\bfseries,
	keywords=[3]{Float, Int, String, T, Void, Bit},
	keywordstyle=[4]\color{orange}\bfseries,
	keywords=[4]{Matrix, Array},
	keywordstyle=[5]\color{darkgreen}\bfseries,
	keywords=[5]{StreamIn, StreamOut, DRAM, ArgIn, ArgOut, HostIO, RegFile, Reg, SRAM, FIFO, LIFO, LUT, LineBuffer},
	keywordstyle=[6]\bfseries,
	keywords=[6]{enq, deq, load, loadTile, mux, store, scatter, gather, :=, value, push, pop, peek},
	keywordstyle=[7]\color{magenta},
	keywords=[7]{until, par, by},
	keywordstyle=\color{magenta}\bfseries,
	morekeywords={Foreach,For,to,for,Reduce,MemReduce,MemFold,Fold,Accel,Stream,FSM,Sequential,if,else,Parallel,Pipe, DummyPipe}
}

\usepackage{multicol}

\usepackage{booktabs}
\newcommand*{\permcomb}[4][0mu]{{{}^{#3}\mkern#1#2_{#4}}}
\newcommand*{\comb}[1][-1mu]{\permcomb[#1]{C}}

\usepackage{tabularx}
\newcolumntype{M}{>{\centering\arraybackslash}m{5.5cm}}
\newcolumntype{L}{>{\centering\arraybackslash}m{1cm}}
\definecolor{vbgray}{gray}{0.9}
\definecolor{darkgreen}{RGB} {0, 100, 0}
\definecolor{darkred}{RGB} {255, 0, 0}

\definecolor{vbgray}{gray}{0.9}
\definecolor{darkgreen}{RGB} {0, 100, 0}
\definecolor{darkred}{RGB} {255, 0, 0}
\definecolor{blue}{RGB} {0, 135, 255}
\definecolor{yellow}{RGB} {224, 173, 0}
\definecolor{codegreen}{RGB}{51,123,0}
\definecolor{codegray}{rgb}{0.5,0.5,0.5}
\definecolor{codepurple}{rgb}{0.58,0,0.82}
\definecolor{backcolour}{rgb}{0.95,0.95,0.95}
\definecolor{lightback}{rgb}{0.95,0.95,0.95}

\usepackage{mathtools}
\DeclarePairedDelimiter\ceil{\lceil}{\rceil}
\DeclarePairedDelimiter\floor{\lfloor}{\rfloor}

\begin{document}

%% Title information
\title{Efficient Memory Partitioning in Software Defined Hardware}         %% [Short Title] is optional;
                                        %% when present, will be used in
                                        %% header instead of Full Title.
%\titlenote{with title note}             %% \titlenote is optional;
                                        %% can be repeated if necessary;
                                        %% contents suppressed with 'anonymous'
%\subtitle{}                     %% \subtitle is optional
%\subtitlenote{with subtitle note}       %% \subtitlenote is optional;
                                        %% can be repeated if necessary;
                                        %% contents suppressed with 'anonymous'

%% Author information
%% Contents and number of authors suppressed with 'anonymous'.
%% Each author should be introduced by \author, followed by
%% \authornote (optional), \orcid (optional), \affiliation, and
%% \email.
%% An author may have multiple affiliations and/or emails; repeat the
%% appropriate command.
%% Many elements are not rendered, but should be provided for metadata
%% extraction tools.

\author{Matthew Feldman}
\affiliation{
  \department{Electrical Engineering}              %% \department is recommended
  \institution{Stanford University}            %% \institution is required
  \country{United States}                    %% \country is recommended
}
\email{mattfel@stanford.edu}          %% \email is recommended

\author{Tian Zhao}
\affiliation{
  \department{Electrical Engineering}             %% \department is recommended
  \institution{Stanford University}           %% \institution is required
  \country{United States}                   %% \country is recommended
}
\email{tianzhao@stanford.edu}         %% \email is recommended

\author{Kunle Olukotun}
\affiliation{
	\department{Electrical Engineering and Computer Science}             %% \department is recommended
	\institution{Stanford University}           %% \institution is required
	\country{United States}                   %% \country is recommended
}
\email{kunle@stanford.edu}         %% \email is recommended

%% Abstract
%% Note: \begin{abstract}...\end{abstract} environment must come
%% before a\maketitle command
\begin{abstract}
As programmers turn to software-defined hardware (SDH) to maintain a high level of productivity while programming hardware to run complex algorithms, heavy-lifting must be done by the compiler to automatically partition on-chip arrays.
In this paper, we introduce an automatic memory partitioning system that can quickly compute more efficient partitioning schemes than prior systems.
Our system employs a variety of resource-saving optimizations and an ML cost model to select the best partitioning scheme from an array of candidates.
We compared our system against various state-of-the-art SDH compilers and FPGAs on a variety of benchmarks and found that our system generates solutions that, on average, consume  40.3\% fewer logic resources, 78.3\% fewer FFs, 54.9\% fewer Block RAMs (BRAMs), and 100\% fewer DSPs.

\end{abstract}

%% 2012 ACM Computing Classification System (CSS) concepts
%% Generate at 'http://dl.acm.org/ccs/ccs.cfm'.

%\ccsdesc[500]{Software and its engineering~General programming languages}
%\ccsdesc[300]{Social and professional topics~History of programming 
%	languages}
%% End of generated code

%% Keywords
%% comma separated list
%\keywords{memory partitioning, cost estimation, loop analysis}  %% \keywords are mandatory in final camera-ready submission

%% \maketitle
%% Note: \maketitle command must come after title commands, author
%% commands, abstract environment, Computing Classification System
%% environment and commands, and keywords command.
\maketitle

\input{text/introduction.tex}
\input{text/background.tex}
\input{text/design.tex}
\input{text/results.tex}

\input{text/relatedwork.tex}

\input{text/conclusion.tex}

%% Acknowledgments
\iffalse
\begin{acks}                            %% acks environment is optional
                                        %% contents suppressed with 'anonymous'
  %% Commands \grantsponsor{<sponsorID>}{<name>}{<url>} and
  %% \grantnum[<url>]{<sponsorID>}{<number>} should be used to
  %% acknowledge financial support and will be used by metadata
  %% extraction tools.
  This material is based upon work supported by the
  \grantsponsor{GS100000001}{National Science
    Foundation}{http://dx.doi.org/10.13039/100000001} under Grant
  No.~\grantnum{GS100000001}{nnnnnnn} and Grant
  No.~\grantnum{GS100000001}{mmmmmmm}.  Any opinions, findings, and
  conclusions or recommendations expressed in this material are those
  of the author and do not necessarily reflect the views of the
  National Science Foundation.
\end{acks}
\fi

%\def\bibfont{\normalsize}

%% Bibliography
\bibliography{./references.bib}
%
%%% Appendix
%\appendix
%\input{text/model_param}
%\input{text/correction}
%%\section{Banking Equations}
%% Text of appendix \ldots

\end{document}

%% file: text/introduction.tex
\section{Introduction}

In recent years, there has been a growing demand for compute devices that can efficiently run increasingly complex algorithms.
Experts from a wide variety of domains, such as Machine Learning \cite{survey}, Computational Physics \cite{physics}, and Genomics \cite{darwin}, are facing unprecedented challenges in processing massive data sets quickly and efficiently.
Spatial architectures, such as field-programmable gate arrays (FPGAs) and reconfigurable dataflow architectures (RDAs), provide performance-per-watt advantages over CPUs and GPUs~\cite{perfperwatt} and are becoming popular for these kinds of domain experts.  
They allow the programmer to create a digital circuit that can act as an accelerator for a particular algorithm.
Hardware accelerators perform computation by flowing data through a pipelined circuit, rather than by serially executing a sequence of instructions. 
Unfortunately, they traditionally require complicated register-transfer level (RTL) languages, like VHDL and Verilog, that domain experts often find prohibitively difficult to use.

For this reason, software-defined hardware (SDH) has emerged as an active research field to solve this problem of programmability without sacrificing performance. 
For certain domains, there are domain specific languages (DSLs), such as Halide \cite{halide}, DNNWeaver \cite{dnnweaver}, Aetherling \cite{aetherling}, RIPL \cite{ripl}, and SODA \cite{soda}, which limit what the programmer can express in order to generate high performance designs for a particular domain.
There are also numerous general purpose SDH languages available today, including Vivado High Level Synthesis (HLS)~\cite{vivadohls}, SDAccel~\cite{sdaccel}, OpenCL~\cite{opencl_sdk}, and Spatial~\cite{spatial}.  
These languages are embedded in high level software languages, such as C or Scala, and allow the programmer to use high level abstractions to create arbitrary hardware accelerators.

SDH languages allow the programmer to exploit both \textit{pipeline} and \textit{spatial} parallelism by nesting loops and annotating iterators.
\textit{Pipeline} parallelism refers to concurrent execution of consecutive stages in a computation graph, and \textit{spatial} parallelism refers to the concurrent execution of the same stage in a computation graph.
These frameworks also allow the programmer to declare multidimensional arrays, or memories, that reside on-chip.
The interplay between parallelism and how these memories are accessed in general-purpose SDH languages is the focus of this paper.
"Memory banking" is the process in which the compiler decides how to manage concurrent accesses to on-chip memories by partitioning them across multiple physical resources.

In the worst case, a banking system that is not well optimized could increase the compile time of an SDH program by minutes to hours, as we show in this paper.
This increase in the programmer's development cycle could significantly hinder the productivity of the SDH tool.
Furthermore, a poorly optimized banking system could also result in wildly inefficient banking schemes.
A naive banking scheme could result in more resources being dedicated to the implementation of the banking scheme than those dedicated to the datapath of the algorithm that does actual computational work.
These two issues could render an SDH framework entirely unusable for a performance-oriented programmer.

State-of-the-art implementations of banking analysis were generally designed for certain kinds of compute patterns in mind, such as linear algebra \cite{spatial} and stencil operations \cite{cong}.  
However, these approaches do not scale when presented with more challenging banking problems that are found in applications with more dynamic access patterns, larger parallelization factors, and more complex control structures.

In this paper, we introduce a novel banking system capable of automatically solving challenging banking problems with solutions that are more efficient than existing state-of-the-art tools.  
Our system makes the following contributions:
\begin{itemize}
	\item Use heuristics to quickly identify a collection of valid banking schemes (Section \ref{search}).
	\item Apply targeted transformations to the datapath associated with each solution. (Section \ref{transform}).
	\item Rapidly estimate resource utilization for each solution using a machine learning (ML) pipeline (Section \ref{cost}).
\end{itemize}

\begin{figure}
	\centering
	\includegraphics[width=0.35\textwidth]{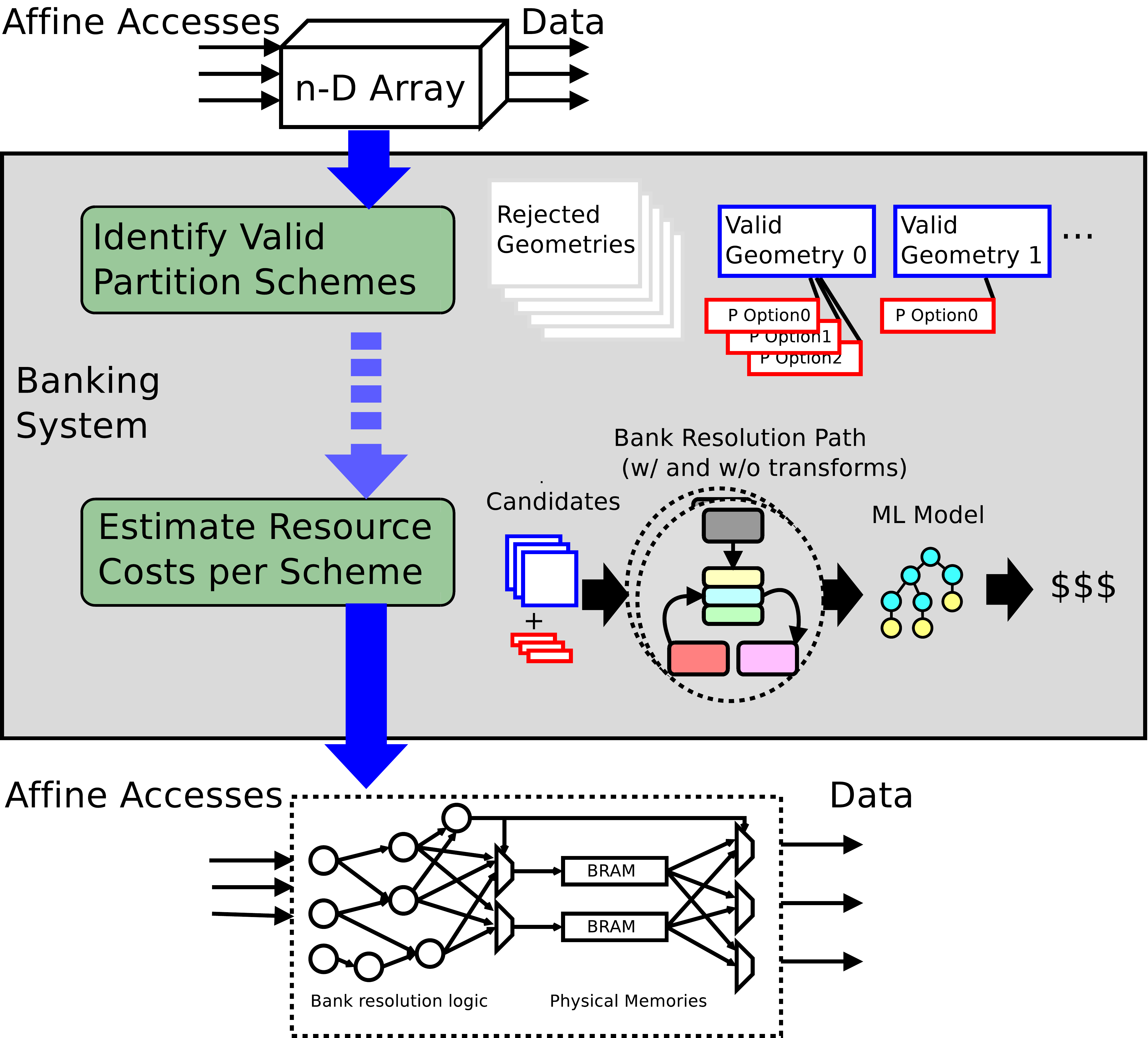}
	\caption{Top level process of banking system.}
	\label{fig:toplevel}
	
\end{figure}

Figure \ref{fig:toplevel} shows the conceptual view of our system.   
The input is logical accesses to an array, coupled with information about the concurrency of these accesses.
The output is an elaborated, retimed circuit that implements memory virtualization to satisfy the constraints of the original program.

We compared our system against various state-of-the-art SDH compilers on a variety of benchmarks and found that our system generates solutions that, on average, consume  40.3\% fewer logic resources, 78.3\% fewer FFs, 54.9\% fewer BRAMs, and 100\% fewer DSPs without losing performance.

%% file: text/background.tex
\section{Background}
\label{background}

\textit{Memory banking} is the process of partitioning a multidimensional array so that it can be implemented in hardware as a collection of physical BRAM resources and surrounding logic to index into them.
In this paper, we refer to the access pattern to an array in the program as the \textit{logical} access (i.e. \texttt{arr[2*i+3]}), and the circuits which connect to BRAMs as \textit{physical} accesses.

\subsection{Representing The Problem Using the Polyhedral Model}
% The \textit{polyhedral model} \cite{memwithpoly} is a formalism for representing a program's iteration space, memory space, and access patterns in a compact way that a compiler can statically analyze.
% We use it as a tool to solve the banking problem, and briefly summarize the important concepts used in prior work \cite{cong}.

The \textit{polyhedral model} is a formalism for representing a program's iteration space, memory space, and access patterns algebraically for a compiler to statically analyze \cite{karp1967organization}.
 It has been used by others to solve various problems such as synthesis, verification, and optimization of systolic arrays, detecting parallelism and efficiently scheduling programs in parallel compute environments, and on-chip memory partitioning in dataflow architectures \cite{feautrier1992somep1, feautrier1992somep2,feautrier1991dataflow, rajopadhye1989synthesizing, rajopadhye1986synthesizing, quinton1989mapping, rajopadhye1986synthesis, memwithpoly, cong, benabderrahmane2010polyhedral, polyhedral}.

 In this work, we use it as a tool to solve the banking problem, and briefly summarize the important concepts used in prior work as it relates to memory partitioning on FPGAs.

\begin{definition}{A \textit{Polyhedron}}
	is a set of points in n-dimensional space that satisfy a set of linear inequalities. Namely, the set of points $P = \{\vec{x} \in \mathbb{Q}^n | A\cdot\vec{x}\leq \vec{b}\}$ where $\mathbb{Q}^n$ is an n-dimensional vector of rational numbers, A is a matrix where $A_{i,j}\in \mathbb{Q}$, and $\vec{b}\in \mathbb{Q}^n$.
\end{definition}
\begin{definition}{A \textit{Polytope}}
	is a bounded polyhedron, e.g. one that contains a finite amount of integer points.
\end{definition}
\begin{definition}{A \textit{Parallelotope}} is a polytope where every pair of opposite facets are congruent and parallel, i.e. the n-dimensional generalization of a parallelogram (2D) or parallelepiped (3D).
\end{definition}
\begin{definition}{The \textit{iterator space}}
	is a polytope defined by the iterators, $\vec{i}$, of a loop nest. 
	%We require that all iterators start at 0 and have a stride of 1. 
	%In the case of an iterator, $k$, in the form \texttt{for (k = B -> S -> E par P)}, we express it in the form $k_\ell = i_\ell*S*P + B + \ell*S$  $\forall i_\ell \in \{0,..,\ceil{\frac{E-\left(B + \ell \cdot S\right)}{S\cdot P}}\}, \ell \in \{0,..,P-1\}$.
	%We refer to $\ell$ as the ``lane."
\end{definition}
\begin{definition}{An \textit{array}}
	is an n-dimensional on-chip memory whose addresses represent a polytope
\end{definition}
\begin{definition}{An \textit{access pattern} for access $a$}
	is an expression representing the mapping from the iterator space to an n-dimensional reference to the array polytope.
\end{definition}
\begin{definition}{An \textit{access group}}
	is a collection of access patterns which can be active during the same cycle at runtime.
\end{definition}

\subsection{Partitioning Methods}

\begin{figure}
	\centering
	\begin{subfigure}{0.23\textwidth}
		\centering
		\includegraphics[width=0.75\textwidth]{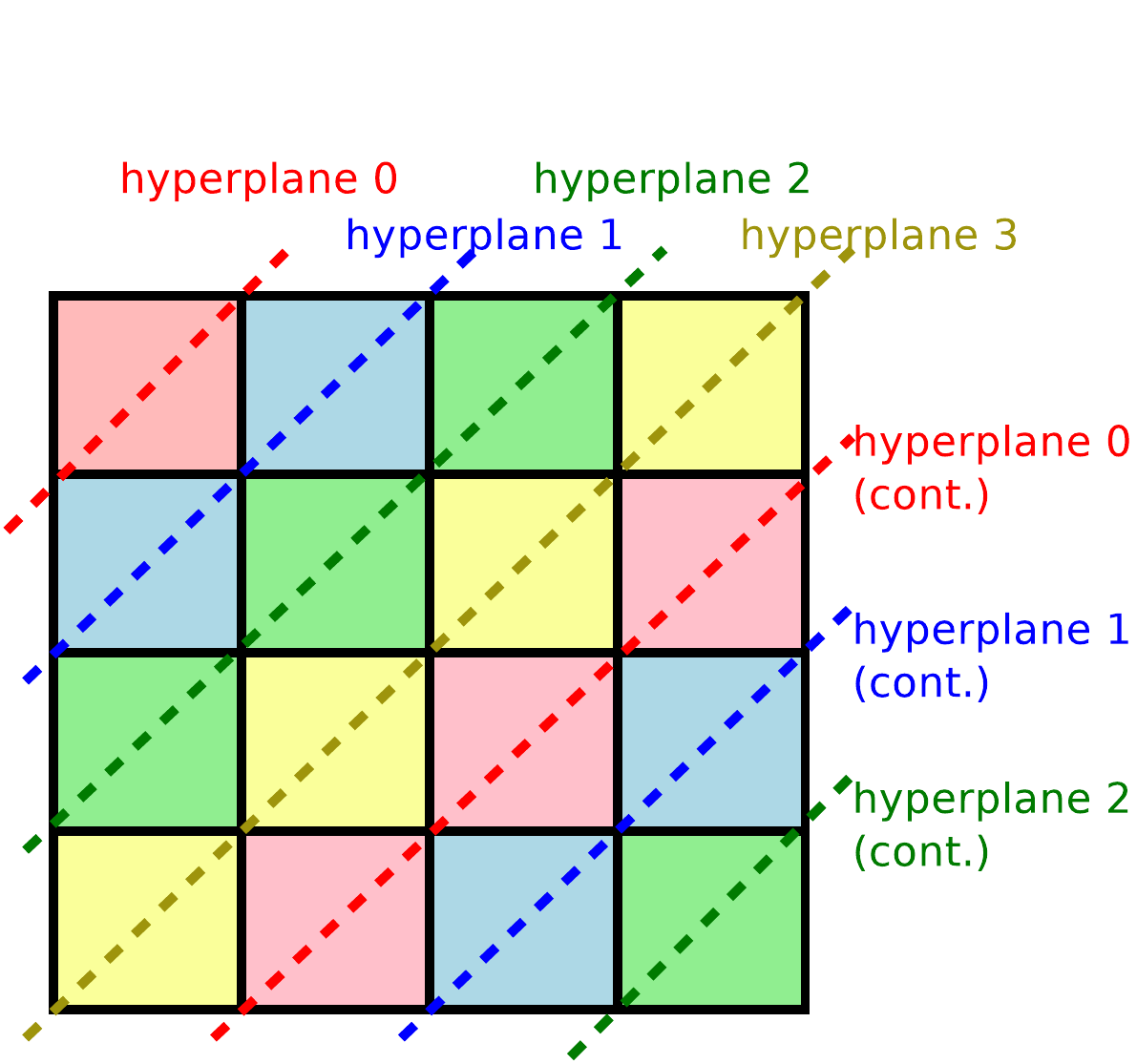}
		\caption{}
		\label{fig:hyperplane}
	\end{subfigure}
	~
	\begin{subfigure}{0.23\textwidth}
		\centering
		\includegraphics[width=0.75\textwidth]{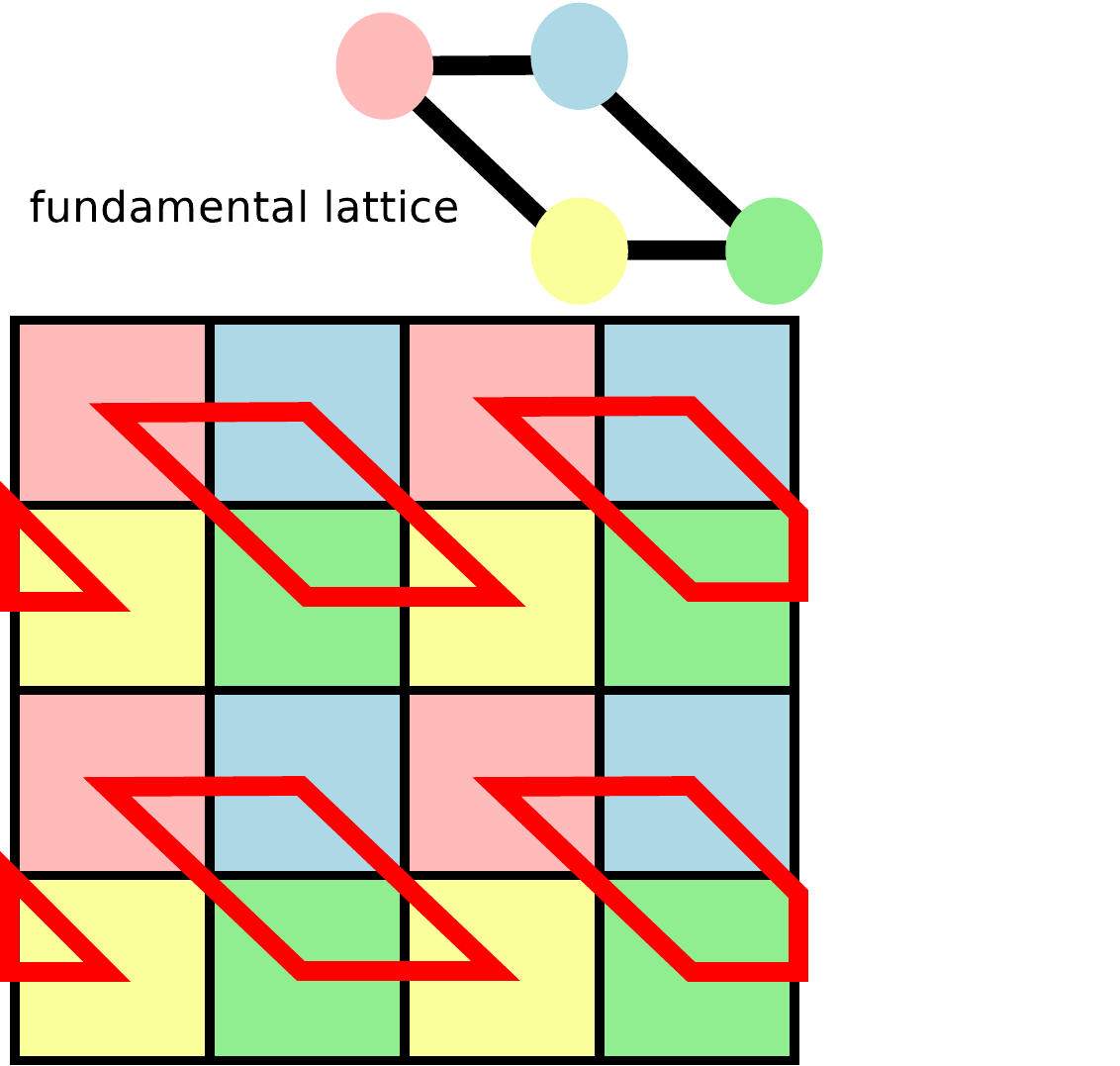}
		\caption{}
		\label{fig:lattice}
	\end{subfigure}
	
	\caption{Example 4x4 array partitioned into 4 banks using (a) hyperplane and (b) lattice.  Color represents bank}
	\label{fig:hyperplanelattice}	
	\vskip-.1in
\end{figure}

There are two common methods for partitioning the memory: \textit{hyperplane partitioning} \cite{cong} and \textit{lattice partitioning} \cite{latticepartition}, sketched in Figure \ref{fig:hyperplanelattice}.
Under hyperplane partitioning, the array polytope is divided into parallel ``hyperplanes," where each hyperplane represents a bank.
Under lattice partitioning, the array is divided into a tessellation of congruent parallelotopes, where each position within the parallelotope is assigned to a unique bank.
While most common partitioning problems have both hyperplane and lattice solutions, there are certain patterns where only one can provide a solution.
For example, hyperplane partitioning can solve problems that require a \textit{block-cyclic} pattern of banks, while lattice partitioning can solve problems with concurrent accesses that do not lie within a rectangular parallelotope

In this paper, we introduce a system based on hyperplane partitioning with an extension that captures a subset of lattice partitioning solutions, namely those composed of orthogonal parallelotopes.  
These are referred to as ``multidimensional" hyperplane geometries and are discussed in \ref{projections}.

Table \ref{table:params} concisely summarizes all of the quantities involved in our banking system.
\textbf{Input parameters} are provided by the SDH framework and represent the access pattern the programmer specified in the code.
\textbf{Solve parameters} are computed by our banking system.
Our system internally computes a collection of these but only returns the optimal set.
\textbf{Metrics} which are helpful quantities for understanding how the partitioning scheme will map to hardware.

\begin{table}
	\begin{tabular}{Llp{5.5cm}}
		\toprule
		\textbf{Type} & \textbf{Name} & \textbf{Description} \\
		\midrule
		\multirow{5}{1cm}{Input} & $\vec{D}$ & Dimensions of memory, $|\vec{D}|=n$ \\
		& $\vec{a}$ & Logical accesses, grouped by concurrency \\
		& $\vec{i}$ & iterators in the scope of an access \\
		& $\vec{x}$ & Memory address reference for an access, $|\vec{x}| = n$      \\
		& $\textit{k}$ & Number of ports on underlying BRAM \\
		\midrule
		\multirow{4}{1cm}{Solve} & $\vec{N}$ & Number of banks, $|\vec{N}|\in\{1,n\}$ \\
		& $\vec{B}$ & Blocking factor, $|\vec{B}| = |\vec{N}|$      \\
		& $\vec{\alpha}$ & Partition vector, $|\vec{\alpha}|=n$    \\
		& $\vec{P}$ & Partition parallelotope,  $|\vec{P}|=n$      \\		\midrule
		\multirow{3}{1cm}{Metrics} & $FO_a$ & \# of banks an access touches (Fan-Out), \\
		& & $FO_a\leq\prod \vec{N}$ \\
		& $FI_b$ & \# of accesses a bank feeds (Fan-In),\\ 
		& & $\sum{FI_b} = \sum{FO_a}$ \\
		& $\vec{\delta}$ & Per-dimension padding\\
		\bottomrule
	\end{tabular}
	\caption{Partitioning parameters and definitions.}
	\label{table:params}
	%	\vskip-0.3in
\end{table}

Equations \ref{eq:BA} and \ref{eq:BO} show how the solve parameters are used to compute a \textbf{bank address ($BA$)} and \textbf{offset ($BO$)}. 
We refer to these collectively as the ``bank resolution" equations.

%\noindent\begin{minipage}[t][][b]{0.45\linewidth}
%$\displaystyle
%BA = \floor{\frac{\vec{x}\cdot\vec{\alpha}}{B}} \bmod{N}
%$
%\begin{equation}\label{eq:BA}\end{equation}
%\end{minipage}\hspace{5px}
%\begin{minipage}{0.52\linewidth}
%$\displaystyle R = \sum_{i=0}^n\left(\floor{\frac{\vec{x}_i}{\vec{P}_i}} \cdot\prod_{j=i+1}^n\ceil{\frac{\vec{D_j}}{\vec{P_j}}}\right)$ \\
%%$\displaystyle C = \sum_{i=0}^{n}{\left(\vec{x_i}\cdot\vec{\alpha_i}\right) \bmod B}$ \\
%$\displaystyle BO = R * B + C$
%\begin{equation}\label{eq:BO}\end{equation}
%
%\end{minipage}

\begin{equation}
\label{eq:BA}
BA = \floor{\frac{\vec{x}\cdot\vec{\alpha}}{B}} \bmod{N}
\end{equation}
\begin{equation}
\label{eq:BO}
BO = \left[B \cdot \sum_{i=0}^n\left(\floor{\frac{\vec{x}_i}{\vec{P}_i}} \cdot\prod_{j=i+1}^n\ceil{\frac{\vec{D_j}}{\vec{P_j}}}\right)\right] + \left(\vec{x}\cdot\vec{\alpha}  \bmod B\right)
\end{equation}

The equation for $BA$ divides the array polytope into parallel hyperplanes, and we therefore refer to $\vec{\alpha}$, $N$, and $B$ as representing a ``hyperplane geometry."
The parameter $\vec{P}$ is only used to compute a physical offset, and a hyperplane geometry may have many valid choices for $\vec{P}$.
It is related to the periodicity of Equation \ref{eq:BA} and represents a region in the array such that each $BA$ appears at least once and no more than $B$ times.

\begin{definition}{A \textit{conflict polytope} is the polytope generated by applying Equation \ref{eq:BA} to the delta between the address patterns of two different accesses ($BA(\vec{x}_{a_1}-\vec{x}_{a_2})$)}
\end{definition}
\begin{definition}{A hyperplane geometry is deemed \textit{valid} for a $k$-ported memory}
	if no set of $k$ accesses contain more than $k-1$ pairs of non-empty conflict polytopes.
\end{definition}

These definitions allow our system to handle addresses with non-affine components using quantifier-free Presburger arithmetic \cite{presberder}.  
Namely, any function call without side-effects (including indirection arrays) that is used in an address expression (i.e. \texttt{f($i_0$)} in \texttt{arr[f($i_0$) + $i_1$]}) can be represented an uninterpreted function symbol.  
This means that even though the compiler cannot analyze $BA($\texttt{f($i_0$) + $i_1$}$)$, it \textit{can} analyze any conflict polytope it generates against another access containing the same function symbol.

Finally, $FO_a$ and $FI_b$ describe the fan-out of accesses and fan-in of banks, respectively. 
They describe the size of the crossbars required to arbitrate between accesses and banks.
If $\vec{P}_i$ does not evenly divide $\vec{D}_i$, then the banking equations result in mathematically ``inaccessible" elements that require the compiler to pad the array with $\vec{\delta}$.
When $B>1$, there may be more inaccessible elements due to unequal representation of each bank within a $\vec{P}$ region.

\subsection{Consequences of Parameters}
\label{consequences}
The values of $\vec{N}$, $\vec{B}$, $\vec{\alpha}$, and $\vec{P}$ have implications on the overall FPGA resource utilization and latency of the circuit, since they impact the bank resolution logic, crossbar sizes, and bank volumes which must all map to quantized resources on the FPGA.

\begin{figure}
	\begin{subfigure}{0.24\textwidth}
		\centering,
		\begin{tabular}{p{0.15\linewidth}}
			\toprule
			{\begin{lstlisting}[language=Spatial,linewidth=0.98\columnwidth, mathescape=true]
for (k = 0->3->M par 2)  
  .. = m[k+1] + m[k+2];
				\end{lstlisting}}\\
			\bottomrule
		\end{tabular}
		\caption{}
		\label{fig:toy}
	\end{subfigure}
	~\hspace{3mm}
	\begin{subfigure}{0.19\textwidth}
		\centering
		\includegraphics[width=\textwidth]{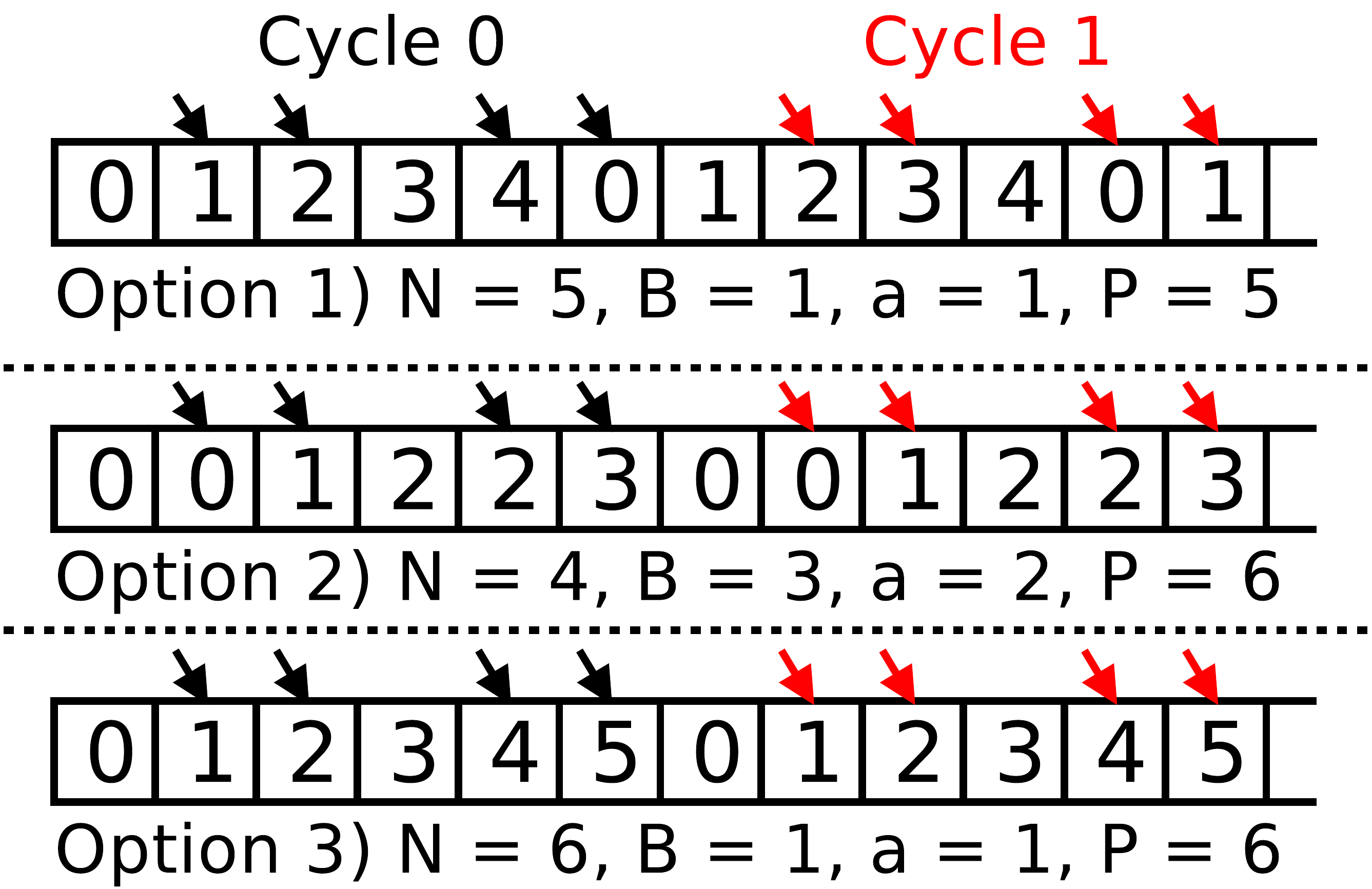}
		\caption{}
		\label{fig:options}	
		%	\begin{subfigure}{1\textwidth}
		%		\centering
		%		\includegraphics[width=\textwidth]{figs/option1-eps-converted-to.pdf}
		%		\caption{}
		%		\label{fig:option1}
		%	\end{subfigure}
		%
		%	\begin{subfigure}{1\textwidth}
		%		\centering
		%		\includegraphics[width=\textwidth]{figs/option2-eps-converted-to.pdf}
		%		\caption{}
		%		\label{fig:option2}
		%	\end{subfigure}
		%
		%	\begin{subfigure}{1\textwidth}
		%		\centering
		%		\includegraphics[width=\textwidth]{figs/option3-eps-converted-to.pdf}
		%		\caption{}
		%		\label{fig:option3}
		%	\end{subfigure}
	\end{subfigure}
	\caption{(a) Sample access pattern with iterator $k$ start=0, step=3, stop=M, and parallelization=2. (b) Three valid banking schemes for \texttt{arr} with $BA$ values shown.}
	\label{fig:sample}	
	\vskip-.1in
\end{figure}

Consider the snippet in Figure \ref{fig:toy} with access patterns $6\cdot i + 1$, $6\cdot i + 2$, $6\cdot i + 4$, and $6\cdot i + 5$ (we substitute $k$ for unit-step iterator $i$).
Figure \ref{fig:options} shows three potential ways this problem can be solved.

Option 1 uses 5 banks and full crossbars for each access ($FO_a$ = 5).
Option 2 appears better since it uses only 4 banks and $FO_a$ = 1, but requires a costly divide-by-3 and a multiply-by-2 operation.  
Option 3 uses 6 banks but achieves $FO_a$ = 1 while also eliminating B and $\alpha$ arithmetic.
Depending on the size of the array and the target FPGA, the volume of one ``bank" may spill into 2 BRAMs, so solutions with smaller $N$ may consume \textit{more} BRAMs.

The key point is that there are always many solutions to a banking problem and it is difficult to predict which one is the best.
Additionally, the definition of ``best" may change depending on which resource is scarcest for a given application.

\subsection{Concepts in Hierarchically-Nested State Machine Programming}
\label{progmodel}
A programming model based on parallel patterns \cite{parallelpatterns} is a good starting point for exposing a high level of abstraction without losing handles on performance \cite {spatial}.
For this reason, we chose to use Spatial as the host compiler for our banking system since it is an open-source, extensible language that can express massive design spaces, an explicit memory hierarchy, and multi-level parallelism.	

Our system targets programs composed as a set of hierarchically nested state machines, or \textit{controllers}.
A \textit{controller} is expressed as a traditional software loop and is represented in the IR as a multi-level counter chain that feeds iterator values to a basic block. 
A \textit{multi-level counter chain} is a counter that spans a multidimensional iteration space whose bounds do not need to be compile-time static.

\subsubsection{Controller Level} 
The contents of a controller's basic block define it's \textit{level}:

\begin{itemize}
	\item \textbf{Inner controllers} only contain dataflow graphs made up of primitives.
	\item \textbf{Outer controllers} only contain other controllers
\end{itemize}

An outer controller and the controllers in its basic block define a \textit{parent-child} relationship in the controller hierarchy.
An outer controller's \textit{width} is the number of children it contains.
A controller's \textit{sub-tree} is the set of controllers found by following its children recursively.
A node's \textit{ancestors} is a list of every controller in the hierarchy that encloses it.
The \textit{least common ancestor} (LCA) of two nodes is the most deeply-nested controller which they both share.

\subsubsection{Controller Schedule}
\label{schedules}
"Scheduling" refers to how nodes in a controller's basic block execute relative to each other but has different meanings for inner and outer controllers.

For outer controllers, there are five schedules describing how the children execute in hardware:
\begin{itemize}
	\item \texttt{Sequential} - Child controllers execute one at a time with no overlap.
	\item \texttt{Pipelined} - Child controllers execute in pipelined  (i.e. overlapping) fashion.
	\item \texttt{Fork-Join} - Child controllers execute simultaneously and independently of each other.  
	\item \texttt{Fork} - One child controller executes per iteration, based on a set of \texttt{if/then/else} conditions.
	\item \texttt{Streaming} - Child controllers execute as long as their input data is available. 
\end{itemize}

For inner controllers, the schedule refers to the mapping from each node in the pipelined dataflow graph to the cycle it will execute.
Scheduling nodes to execute during different cycles allows the design to safely run at a higher clock rate.
Their runtime is entirely defined by their \textit{latency}, which is the cycle at which its latest node is scheduled, and its \textit{initiation interval}, which is the number of cycles the controller must wait before it can increment the multi-level counter and issue another iteration.
The hardware-complexity of the datapath determines the latency, and the loop-carry dependencies determine the initiation interval.

\subsubsection{Controller Parallelization}
\label{pommop}
Parallelization is one of the ways that programs mapped to FPGA can improve performance.
Specifically, parallelizing a loop by $P$ means that $P$ consecutive iterations of the loop will be executed simultaneously.
To a first-order approximation, this means that parallelization results in a factor of $P$ performance improvement by using $P$-times as many resources compared to the un-parallelized loop.

In reality, this typically improves performance by less than $P$ and increases resource utilization by more than $P$.
This is because initiation interval and latency may change with parallelization.
We must distinguish between parallelization of inner and outer controllers separately to understand why.

We walk through the example snippet in Figure \ref{fig:parexample} to demonstrate this.

Parallelization applied to an inner controller results in \textit{vectorization} of the datapath.  
Figure \ref{fig:innervec} shows how this is done for loop \texttt{k} with \texttt{IP} = 2.

Parallelization applied to an outer controller results in \textit{unrolling}.
Each child controller is cloned in whole and added to the hierarchy.
This is shown in Figure \ref{fig:pommop} as the "Pre-Unrolled" tree structure is transformed into the "Intermediate" structure for \texttt{OP} = 2.
The compiler must then inject \texttt{Fork-Join} controllers to this intermediate structure to achieve the parallelism specified in the program.
There are two strategies that the compiler may employ to capture this level of parallelism, shown in the "Post-Unrolled" trees in Figure \ref{fig:pommop}:
\begin{itemize}
	\item \textit{ForkJoin-of-Pipelines unrolling} is when the \texttt{Fork-Join} controllers are injected \textit{between} the outer controller and child stages such that all lanes of each child are synchronized (i.e. always begin their execution simultaneously). This guarantees that child \texttt{c} will not begin executing iteration \texttt{i+P} until iterations \texttt{i/P*P} to \textit{(i/P*P) + P - 1} are completed, for any child and iteration of the parent controller.
	\item \textit{Pipeline-of-ForkJoins unrolling} is when each lane of the original outer is separated into its own controller that is structurally identical to the pre-unrolled loop. A single \texttt{Fork-Join} controller is injected \textit{above} the outer controllers, such that all lanes begin executing simultaneously.  This guarantees that child \texttt{c} will not begin executing iteration \texttt{i+P} until iteration \texttt{i} is completed, for any child and iteration of the parent controller.  
	However, there is no guarantee about the relative ordering between iteration \texttt{i/P*P} and \texttt{i/P*P+K}, where \texttt{K \% P != 0}.
\end{itemize} 

For any node in the IR, its "unroll ID" (UID) is a list of integers describing which lane of each ancestor controller it belongs to.
The UID of a node is described as the "base UID" if all integers in the UID are 0.

\begin{figure}
	\centering
	\begin{tabular}{p{\linewidth}}
		\toprule
		{\begin{lstlisting}[language=Spatial,linewidth=0.98\columnwidth, mathescape=true]
// Parent (Outer Ctrl)
Foreach(N by 1 par OP){i =>  
  // Child 0
  Foreach(M by 1){j => /*...*/}
  // Child 1 (Inner Ctrl)
  Foreach(K by 1 par IP){k => 
    val t = i*3
    mem(k) = mux(k == 0, t + mem(k), t)
  }
}
			\end{lstlisting}}\\
		\bottomrule
	\end{tabular}
	\vspace{-5pt}
	\caption{Example construction of an outer controller (Loop \texttt{i}) with parallelization (\texttt{OP}) and two child controllers (Loops \texttt{j} and \texttt{k}).  Loop \texttt{k} has parallelization factor \texttt{IP}.}
	\label{fig:parexample}
	
\end{figure}

\begin{figure}
	\centering
	\includegraphics[width=0.47\textwidth]{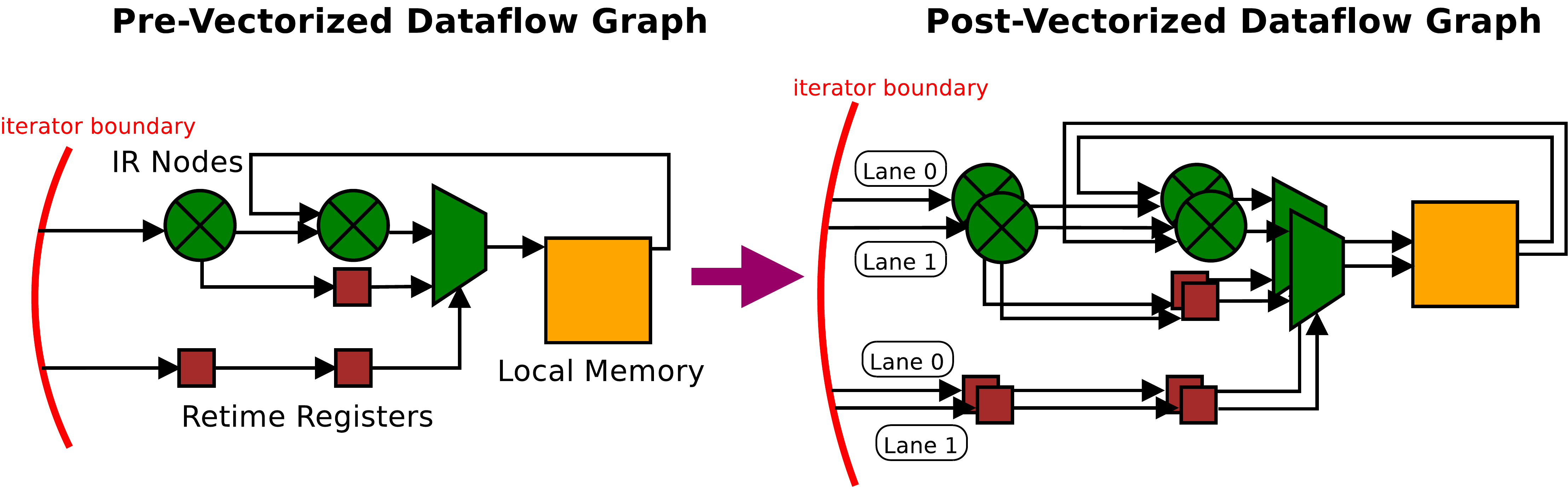}
	\caption{Dataflow path for the inner controller of loop \texttt{k} in Figure \ref{fig:parexample} before and after applying vectorization (IP=2).}
	\label{fig:innervec}
	
\end{figure}
\begin{figure}
	\centering
	\includegraphics[width=0.47\textwidth]{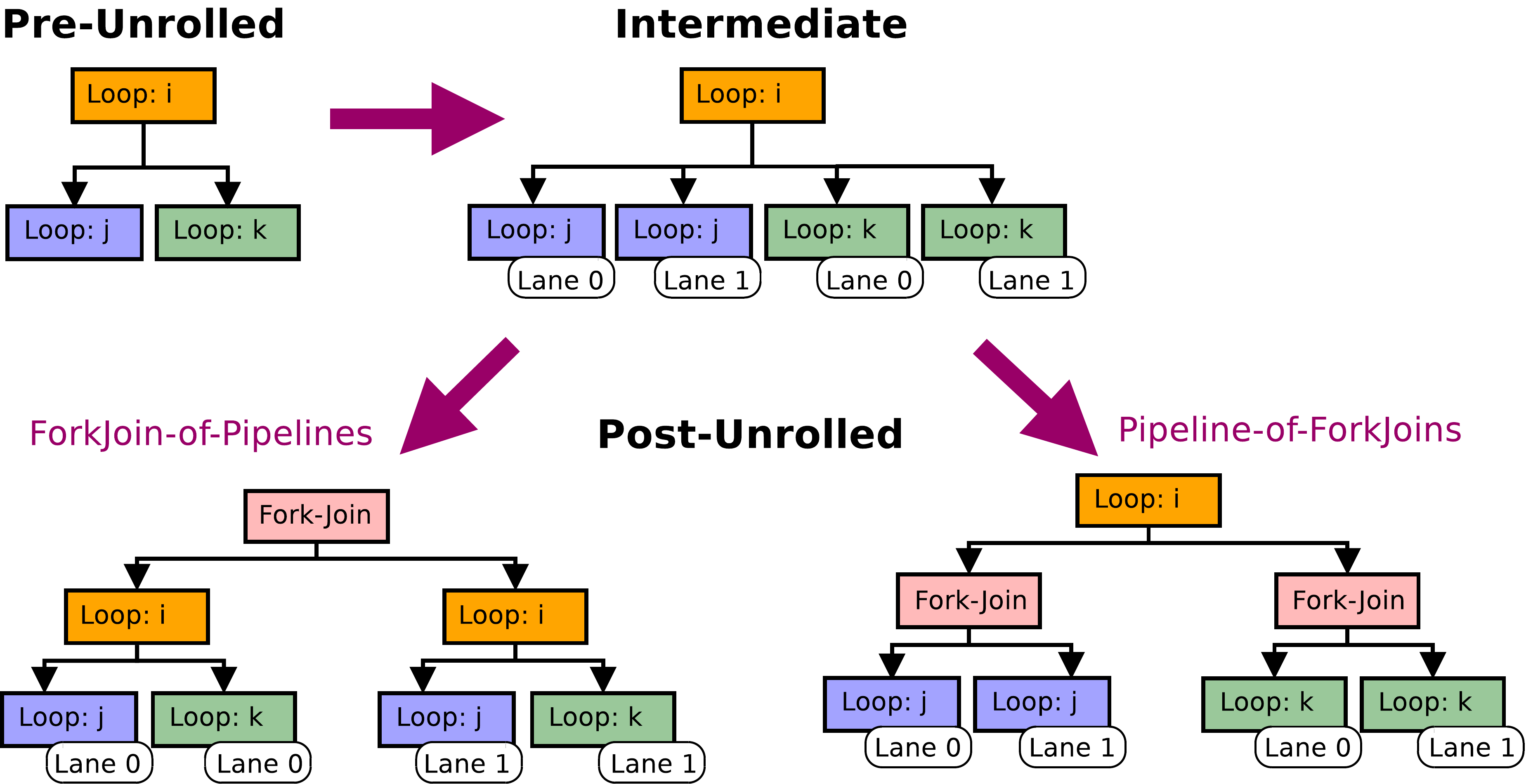}
	\caption{Controller trees before and after unrolling loop i in Figure \ref{fig:parexample} (OP=2, IP=2) for stage-based and lane-based synchronization.}
	\label{fig:pommop}
	
\end{figure}

%% file: text/design.tex
\section{Banking System Design}
\label{system}
In this section, we begin by discussing the analysis that analyzes accesses in a program and converts them into a polytope emptiness problem. 
Then, we describe the three steps in our system for solving this problem:

\begin{itemize}
	\item Compute a list of solutions (constants for $\vec{N}$, $\vec{B}$, $\vec{\alpha}$, and $\vec{P}$) that satisfy the constraints
	\item Apply resource-saving optimizations to the bank resolution datapaths
	\item Estimate the cost of each solution and return the best
\end{itemize}

\subsection{A Running Example}
\label{example} 
\label{sec:example}
In order to drive the components of our banking system, we introduce a motivating example that presents an interesting banking problem.
Figure \ref{fig:mdcode} shows part of the grid-variant of the Molecular Dynamics (MD) benchmark from Machsuite, as it relates to the accesses to one memory. 

At a high level, this algorithm models a 3-dimensional field of molecules.
The molecules are grouped into "cells," which means the field is represented by the four-dimensional structure \texttt{dvec\_sram}. 
The first three dimensions describe the location of a cell in space, and the fourth dimension enumerates each molecule within the cell.
The algorithm computes how each molecule interacts with each molecule in its 26 adjacent cells.

In this snippet, we are initializing the memory in Line 1 by fetching data from DRAM to it and writing \texttt{PL} elements per cycle in its leading dimension.
Later, in Line 7, we read this data from the memory to drive the algorithm.
Note the innermost controller's counter spans \texttt{Q\_RNG(x,y,z)}, which refers to the amount of molecules in the cell at x, y, and z. 
Each cell contains a different number of molecules, so this is a data-dependent value.
There are a total of \texttt{PL} writers and \texttt{PX*PY*PZ*PP*PQ} readers.

\begin{figure}
	\centering
	\begin{tabular}{p{\linewidth}}
		\toprule
		{\begin{lstlisting}[language=Spatial,linewidth=0.98\columnwidth, mathescape=true]
dvec_sram loadTile dvec_dram par PL
Foreach(X_RNG, Y_RNG, Z_RNG par PX,PY,PZ)
  { (x, y, z) =>
    Foreach(P_RNG par PP) { p =>
      Foreach(Q_RNG(x,y,z) par PQ) { q =>
        ... = dvec_sram(x,y,z,q)
      }
    }
    Foreach(P_RNG_dyn to density) { ... }
  }
			
			\end{lstlisting}}\\
		\bottomrule
	\end{tabular}
	\vspace{-5pt}
	\caption{Access pattern on 4D \texttt{dvec\_sram} from the MD\_Grid benchmark.}
	\label{fig:mdcode}
	
\end{figure}

\input{text/synchronization}

\input{text/search}
\input{text/cost}

%% file: text/synchronization.tex
\subsection{Distilling the Program to a Polytope Emptiness Problem}
\label{synchronization}

There are three steps involved in constructing the polytope emptiness problem: group placement, address pattern extraction, and synchronization analysis.

\paragraph{Group placement} is the first step of this process.
A \textit{group} is a collection of accesses which may be active simultaneously on the same \textit{buffer} of a memory.
Each access, $a$, is assigned to a group, $G$, during this step.
If accesses occur in different stages of an outer pipeline controller, they are not grouped together because they would access different buffers.
The banking system only needs to compute a banking scheme that satisfies each group in isolation since only one group will be active at a time.

For the running example, we start by collecting all of our accesses and initializing our groups list, $\vec{G}$ with the first access:

\begin{equation} \begin{split}
\vec{a} &= \{w_1, ..., w_{PL-1}, r_{000}, r_{001}, ..., r_{PX*PY*PZ-1,PP-1,PQ-1}\} \\
\vec{G} &= \{ \{w_0\} \}
\end{split}
\label{eq:enumeration}
\end{equation}

The subscript integers refer to the UID of each access.
$w_n$ refers to vectorization from \texttt{PL} and $r_{ijk}$ refers to the lanes of loops \texttt{x/y/z} (flattened), \texttt{p}, and \texttt{q}.  

The compiler must inspect the ancestors of each access and place it into one group.
The pseudo-code in Figure \ref{alg:grouping} shows how to perform the grouping.
\texttt{isConcurrent} is defined for each controller based on the following:
\begin{itemize}
	\item \textbf{If \texttt{lca} is an inner controller}, \texttt{isConcurrent} returns true if the distance (in cycles) between the scheduling of \texttt{a} and \texttt{b} is less than the initiation interval of the \texttt{lca}.
	\item \textbf{If \texttt{lca} is an outer controller}, \texttt{isConcurrent} returns true if the schedule is Fork-Join or Stream and returns false if the schedule is Sequential, Fork, or Pipelined scheduling.  Note in the case of a Pipelined controller, the accesses are concurrent but routed to different buffers.
\end{itemize}

\begin{figure}
	\begin{tabular}{p{\linewidth}}
		\toprule
		{\begin{lstlisting}[language=Pseudo,linewidth=0.4\paperwidth, mathescape=true]
for a in $\vec{a}$:
	clash = false
	gId = -1
	while !clash && gId < $\vec{G}$.len:
		gId++
		G = $\vec{G}_{gId}$
		for b in G:
			clash = clash || lca(a,b).isConcurrent
		
	$\vec{G}_{gId}$.append(a)
	
			\end{lstlisting}}\\
		\bottomrule
	\end{tabular}
	\vspace{-10pt}
	\caption{Algorithm for computing iterator synchronization.
		\vspace{-15pt}
	}
	\label{alg:grouping}
\end{figure}

The \texttt{lca} of two accesses is often a \texttt{Fork-Join} controller due to the unrolling strategies described in Section \ref{pommop}.  
Applying the algorithm to our running example yields a banking problem with two groups:

\begin{equation} \begin{split}
\vec{G} = \{ &\{w_0, w_1, ..., w_{PL-1}\}, \\
 &\{r_{000}, r_{001}, ..., r_{PX*PY*PZ-1,PP-1,PQ-1}\} \}
\end{split}
\label{eq:solvedgrouping}
\end{equation}

\paragraph{Address Pattern Extraction} is the second step of the process.
In this step, we convert each access into a polytope so that we can apply polytope-emptiness testing to each access conflict within a group \cite{cong}.
An affine address equation is one in the form of:

\begin{equation}
\vec{x} = A_{n\times m}\cdot\vec{i} + C_{n\times 1}
\label{eq:affine}
\end{equation}
  
The accesses in our example can conveniently be written with two parameterized patterns:

\begin{figure}[h!]
\begin{tabular}{c 	c }
	$w_n$ & $r_{ijk}$ \\
	$\left[\begin{array}{cccc|c}
	d0 & d1 & d2 & d3 & c \\
	1 & 0 & 0 & 0  & 0 \\
	0 & 1 & 0 & 0  & 0 \\
	0 & 0 & 1 & 0  & 0 \\
	0 & 0 & 0 & PL  & n \\
	\end{array}\right]$&
	$\left[\begin{array}{cccc|c}
	x & y & z & q & c \\
	PX & 0 & 0 & 0  & i_X \\
	0 & PY & 0 & 0  & i_Y \\
	0 & 0 & PZ & 0  & i_Z \\
	0 & 0 & 0 & PQ  & k \\
	\end{array}\right]$
	\vspace{.07in}\\
	$\begin{bmatrix}
	d0 = \text{0 to W by 1} \\
	d1 = \text{0 to W by 1} \\
	d2 = \text{0 to W by 1} \\
	d3 = \text{0 to N by PL}
	\end{bmatrix}$ &
	$\begin{bmatrix}
	x = \text{X\_RNG by PX} \\
	y = \text{Y\_RNG by PY} \\
	z = \text{Z\_RNG by PZ} \\
	q = \text{Q\_RNG(x,y,z) by PQ}
	\end{bmatrix}$ 
\end{tabular}
\caption{Affine address patterns and iterator constraints for $w_n$ and $r_{ijk}$.}
\label{affineex}
\vskip-.1in
\end{figure}

The iterators d0-3 are implicit in the expansion of the \texttt{loadTile} syntax.
Address pattern expansion can also be applied to accesses that are only partially affine.  
For example, \texttt{... = mem(i + $\delta_0$*j + $\delta_1$)} would treat the \texttt{$\delta_0$*j} as an unbounded iterator and $\delta_1$ as a static iterator with unity range if it were run-time static.

\paragraph{Synchronization} is the final step of this process.
While it may appear that we are ready to apply polytope-emptiness testing to our equations, the solutions we find would almost certainly result in run-time bugs due to bank collisions.
This is because we did not account for iterator synchronization in the SDH programming model.

An iterator from two different UIDs is \textit{synchronized} if it is guaranteed to always increment during the same cycle for both UIDs, start from the same value, and step by the same amount.
The iterator is \textit{partially synchronized} if it is guaranteed to increment during the same cycle for both UIDs, but the starts or steps for the two UIDs always varies by a fixed amount.
The iterator is \textit{unsynchronized} if none of these guarantees can be proven by analyzing the control structures.

The compiler must determine global substitution rules for each iterator based on its UID and apply them uniformly to all accesses so that we achieve a globally-synchronized banking problem.

In our example, \texttt{Q\_RNG(x,y,z)} varies with the first integer in an access' UID.
For example, if PX = 2, then readers $r_{0**}$ and $r_{1**}$ exist in separate subtrees.
Each subtree's loop \texttt{q} experiences a different value for \texttt{Q\_RNG(x,y,z)}, which in turn means that these subtrees will have different execution times.

Under PoF unrolling, all UIDs of loop \texttt{q} will \textit{start} simultaneously but may start from a different value depending on \texttt{Q\_RNG(x,y,z)}.
This means \texttt{q} is unsynchronized between UIDs.
Under FoP unrolling, different UIDs of loop \texttt{q} may initiate at different points in time since different lanes of loop \textit{x/y/z} run independently of one another without synchronization.
\texttt{q}, in addition to \texttt{p} (not used in the address pattern), \texttt{x}, \texttt{y}, and \texttt{z}, would all be unsynchronized between UIDs.

Note that parallelization of iterators that are part of an access' ancestors but not directly used in the address pattern can still impact synchronization.
In this example, we would have to carry out synchronization analysis if \texttt{PP} > 1.

%% file: text/search.tex
\subsection{Building a Solution Set}
\label{search}

Our solution set is a collection of $\vec{N}$, $\vec{B}$, $\vec{\alpha}$, and $\vec{P}$ tuples which satisfy the banking constraints.
Finding this set requires us to build candidate sets for $\vec{N}$, $\vec{B}$, and $\vec{\alpha}$, and check for combinations that can be proven to be valid geometries for the given access pattern.
Then various possible values for $\vec{P}$ are calculated for each geometry.

\paragraph{Prioritizing Candidate Sets}
In order to increase the probability of finding a ``good" solution quickly, we prioritize certain parameters in these candidate sets.
Specifically, we find the LCM of the access groups' sizes and prioritize the first few multiples of this. 
This is more likely to find schemes that do not require full cross-bars between the accesses and the banks (e.g. small $FO_a$).  
We also remove candidates for $\vec{\alpha}$ if the elements are not mutually co-prime with the element(s) in $\vec{B}$, since the same geometry can be expressed by dividing all parameters by the GCD.
Finally, we prioritize integers for all parameters that can be broken down by rules introduced in \ref{transform}.
All other values are de-prioritized.

\paragraph{Multidimensional Banking} 
\label{projections}
In addition to flat schemes \cite{cong}, our system also searches for multidimensional schemes.
In these schemes, we bank each dimension of the memory separately with a 1-dimensional hyperplane geometry based on the \textit{projections} of the original accesses.
This means that we produce a $BA$ per-dimension ($BO$ is still a scalar, namely the intra-bank offset of a multidimensionally-indexed bank)
Multidimensional schemes is the subset of lattice partitioning schemes based on orthogonal parallelotopes.

Multidimensional banking schemes can also be verified more quickly.
This is because $\comb{\ell}{k}$ polytope-emptiness checks must occur to prove $\ell$ concurrent accesses are properly banked for a $k$-ported memory.
The complexity of verification is therefore $\mathcal{O}(\ell^k)$, since $k$ is typically much smaller than $\ell$.
Projecting accesses results in smaller groups per dimension, either due to \textit{redundancy} (two projections are identical) or \textit{regrouping} (two accesses are guaranteed to never conflict because $BA$ on at least one other dimension always differs).
%These cases are shown in Figure \ref{fig:nonconflict}.
This reduction is most significant when the accesses are heavily-parallelized on multiple dimensions, allowing the system to verify banking solutions more rapidly.
%
%\begin{figure}
%		\centering
%		\includegraphics[width=0.25\textwidth]{figs/nonconflict-eps-converted-to.pdf}
%	\caption{Example \textit{redundancy} and \textit{regrouping} drop-out.}
%			\label{fig:nonconflict}
%			\vskip-1em
%	
%\end{figure}
%
%When solving for a multidimensional scheme on a $k$-ported memory, our system only treats \textit{one} dimension as the $k$-ported banking problem, and all other as single-ported.  
%This is because $\prod{\vec{N}} >= \ceil{\frac{\ell}{k}}$ must hold true for each of $\ell$ concurrent accesses to have no conflicts.
%Our system produces a solution for each dimension acting as the $k$-ported one, since it is not always obvious which one is optimal.
%While not strictly necessary, this is convenient for $k\leq2$.

\paragraph{Fewer-Ported Solutions}
If the underlying BRAM supports $k$ ports, there may be some area overhead associated with the memory template when there are more than $k$ accesses that may connect to one bank.
For this reason, our analyzer adds sub-$k$-ported solutions to the solution set.

\paragraph{Bank-by-duplication} 
Our system iteratively partitions readers into separate groups and routes each group to a different duplicate of the array in certain cases.
It then re-runs the banking analysis on each duplicate separately.
Banking-by-duplication is occasionally the best strategy in cases where LUTs are scarce but BRAMs are abundant.

\subsection{Resource-Saving Datapath Transforms} 
\label{transform}
The banking resolution logic (Equations \ref{eq:BA} and \ref{eq:BO}) include multiplication, division, and modulo operations that may be costly on an FPGA.
Because our banking analyzer has the freedom to choose the actual constants used in these equations, we can aim for those constants that allow for resource-saving transformations and avoid calling vendor-specific arithmetic IPs.

\paragraph{Crandall's Algorithm} We apply Crandall's algorithm to perform division and modulo by \textit{Mersenne} numbers (i.e. those in the form $M = 2^n-1$).
This allows us to perform a cascade of bit-wise operations and simple additions rather than calling division or modulo IPs in hardware.
We further rewrite modulo operations when the operand can evenly divide a Mersenne number. 
Specifically, if $M_2 \cdot k = 2^n-1$ for some $1 < k < R$, then we can apply Crandall's algorithm on $M$ followed by a $k$-wide one-hot mux to compute $\mod M_2$.

\begin{equation}
x \mod M_2 \equiv (x \mod M) \mod M_2
\end{equation}

For reference, there are 5 Mersenne integers, 5 integers that evenly divide a Mersenne integers with R=16, and 6 power-of-2 integers between 1 and 65. 
This provides a sufficiently large pool of desirable constants towards which we can steer our system's search.

\paragraph{Binary decomposition of multiplication} We also optimize multiplication in the form of $a*c$, when $c$ is a constant in the form $c = \sum_{0\leq k < R }S(k)2^{n_k}$, where
$S(k) \in \{\pm1\}$ and $R$ is the radius of the optimization.
This allows the compiler to apply the rewrite rule $a*c = \sum_ka*S(k)2^{n_k}$, which is simply the sum of bit-shifts of $a$.

For reference, with $R$=2, half of the integers between 1 and 65 can be rewritten using only bit-shifts and addition.

%% file: text/cost.tex
\subsection{Machine Learning Model for Resources Estimation}
\label{cost}
Finally, the compiler chooses a valid banking scheme requiring minimal resources, 
    i.e., the RTL code generated from the scheme leads to the least hardware resource after the downstream FPGA toolchain completes PnR.
Previous work \cite{cong} used an analytical model to estimate the hardware resource. 
However, we found that building an accurate analytical model is very challenging due to the complexity of our RTL templates.
Instead, we built a machine learning pipeline to provide the searching process with reasonable estimates of the hardware resources required for a banking scheme. 

\subsubsection{Architecture of Resource Estimator}

\begin{figure}
	\centering
    \includegraphics[width=0.4\paperwidth]{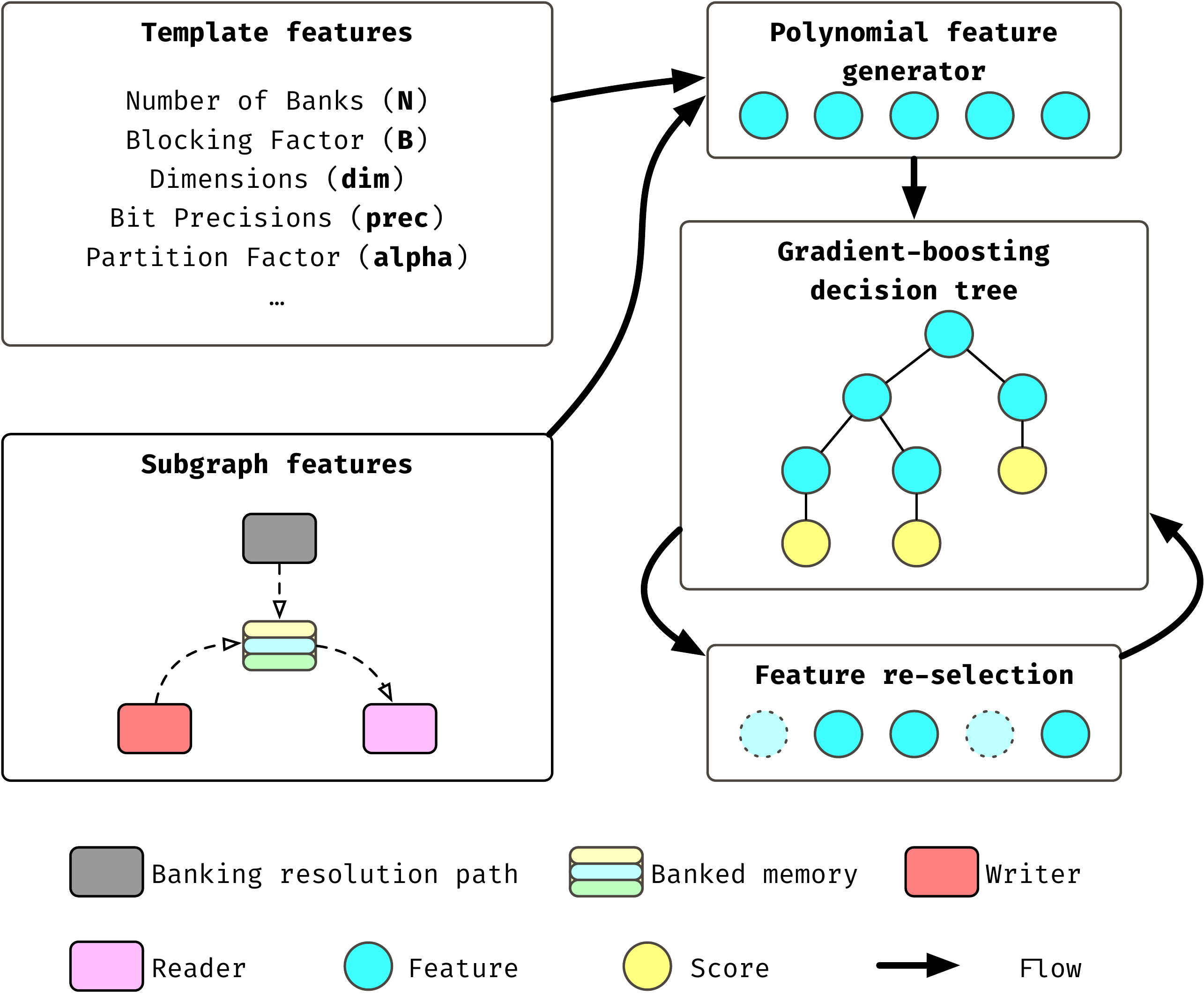}
	\caption{A machine learning architecture to predict a banking scheme's resources after PnR.}
    \label{fig:ml_pipeline}
\end{figure}

Previous work \cite{dse_koeplinger} used a Multi-layer Perceptron (MLP) model for hardware resource estimation.
From our experiments, we found that MLP performed worse on small datasets due to overfitting.
Hence, we built a pipeline based on decision trees and fine-tuned it to reduce overfitting.
Figure \ref{fig:ml_pipeline} shows the pipeline's architecture. 
It takes two classes of features from the parameters of a banking scheme: 
\begin{itemize}
    \item \textbf{Template features} that include primitives and derived parameters.
    \item \textbf{Subgraph features} that include neighbors and accessors of a memory node in the dataflow.
\end{itemize}
The first stage in the pipeline generates second-degree polynomial combinations of raw features.
This approach helps create stronger features, 
    e.g., the product of the number of banks over all the dimensions in a high-dimensional memory node,
    at the expense of generating an ample feature space that can hurt the training speed.
The second stage consists of a regressor based on the gradient-boosting tree \cite{chen2016xgboost}. 
The last stage re-selects the generated features based on their importance.
We define importance as the frequency each generated feature appears in the trained model.
In our experiment, we found that 36 generated features provided the searching process with enough accuracy.

\subsubsection{Training and Fine-Tuning the Estimator}
We created the dataset by using Spatial's regression benchmark suite \footnote{\texttt{Spatial regression suite}: \url{https://github.com/stanford-ppl/spatial/tree/master/test/spatial/tests}}. 
However, our approach can be applied to any SDH framework with a reasonably explicit representation of the memory template and bank resolution logic exposed in its IR.
We ran PnR on all the RTL files generated by this benchmark suite to collect the resources used for every memory and arithmetic node in each application.
A sample in the dataset contains a memory node's raw features and its resources in terms of look-up tables (LUT), flip-flops (FF), and RAMs.
The created dataset contains 831 samples.

Due to the small size of this dataset, we carefully fine-tuned our proposed pipeline to control overfitting.
We trained two models: a baseline MLP model similar to the one proposed in \cite{dse_koeplinger}, and the model pipeline proposed in this work.
We were not able to achieve high performance by using the original baseline model; 
    hence, we augmented its architecture and fine-tuned it to get the best possible performance on the dataset.
Specifically, we performed an exhaustive grid search on the training and regularization parameters for the baseline model and chose the one with the best performance.
Due to the complexity of the proposed model, we were not able to search for the best parameters exhaustively.
Hence, we focused on tuning the regularization parameters to avoid overfitting.
Please refer to Table 3 in the appendix for the final parameters of each model.

\begin{figure}[!tbp]
	\centering
	\begin{subfigure}[b]{0.235\textwidth}
		\includegraphics[width=\textwidth, height=1.2in]{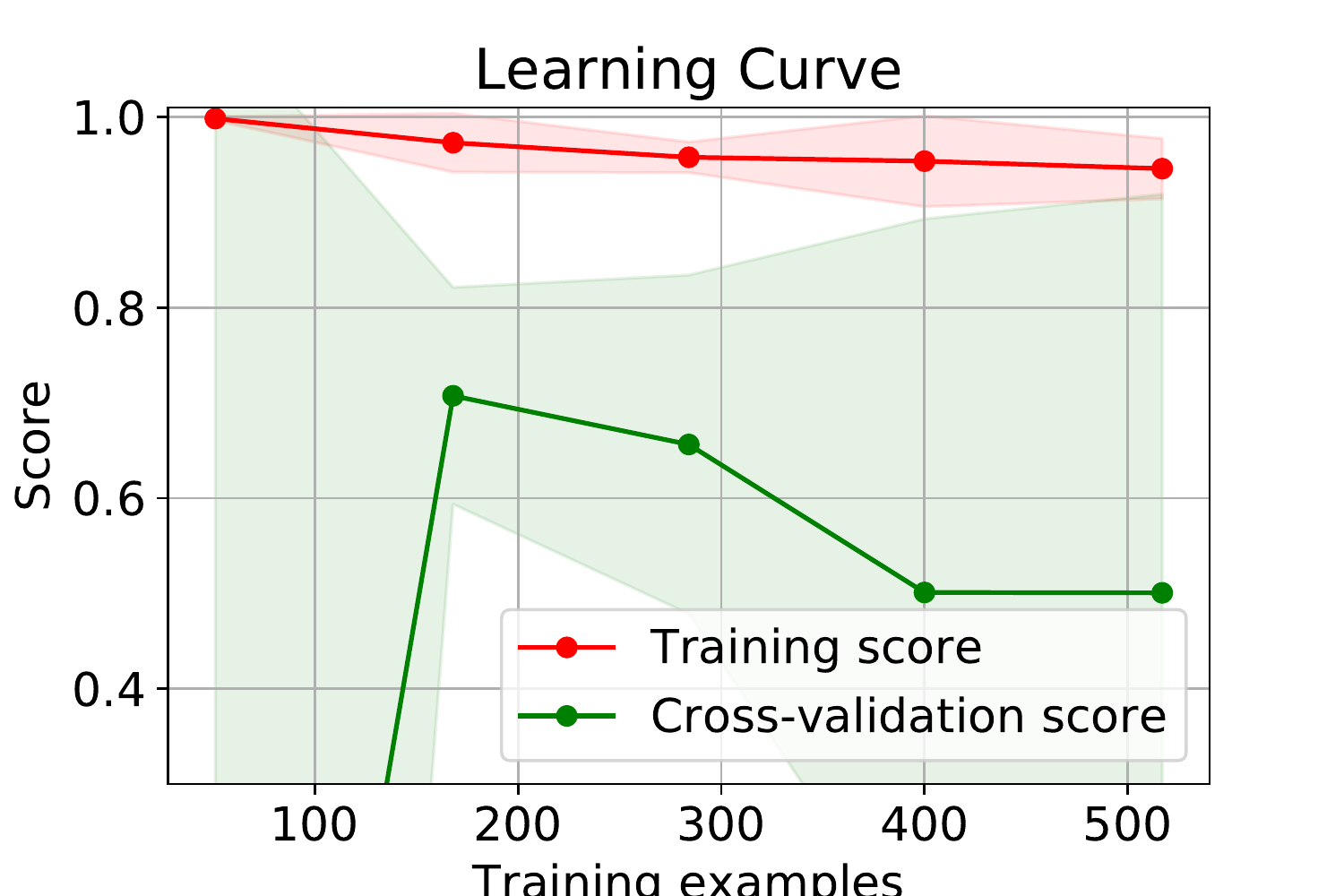}
		\caption{The baseline model.}
	\end{subfigure}
	\begin{subfigure}[b]{0.235\textwidth}
		\includegraphics[width=\textwidth, height=1.2in]{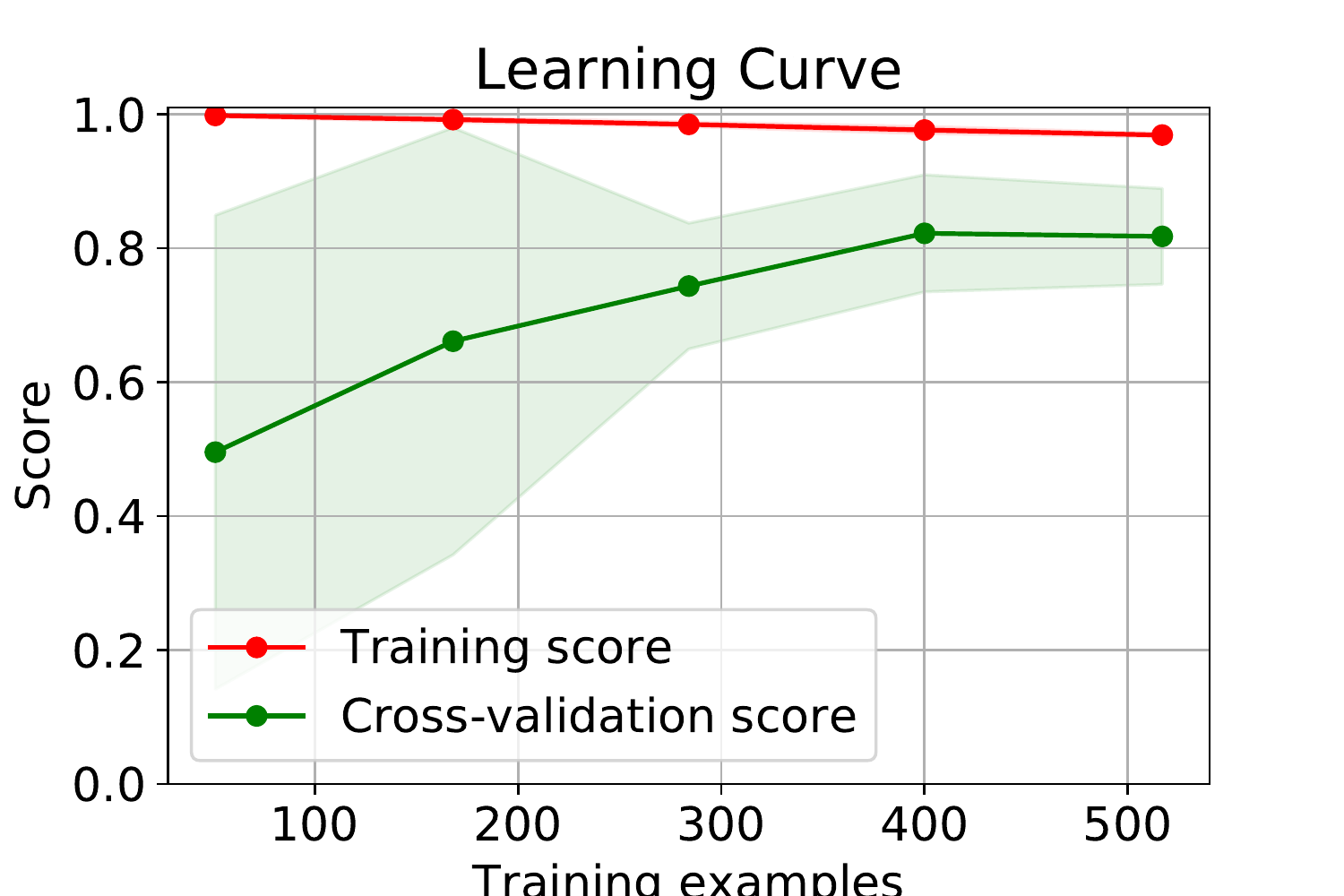}
		\caption{The proposed pipeline.}
	\end{subfigure}
	\caption{Learning curves of the baseline model (left) and our proposed model pipeline when fine-tuned on the dataset from Spatial's regression suite.}
	\label{fig:learning_curves}
\end{figure}

For cross-validation, we randomly permuted the raw dataset ten times.
Every time, we split the dataset into a training set and a test set randomly with a 7-to-3 ratio. 
We collected the training scores and test scores for both models. 
Figure \ref{fig:learning_curves} shows learning curves for both models when predicting the LUT resource.
For learning curves showing both models predicting other resources, please refer to Appendix A.
%\ref{model_param}

In our experiment, we scored both curves using $R^2$. 
We averaged the score curves over all the ten splits.
The colored fields show the standard deviation of all the scores collected during cross-validation.
Our proposed model pipeline achieves an average $R^2$ score of 0.86, which is higher than the baseline model's average $R^2$ score of 0.60.
Besides, the model shows a smaller standard deviation of its cross-validation test scores, which indicates that it suffers less from overfitting.
%TODO: Do I have to report the MAE? 
% The average mean absolute error (MAE) of $186.68$. The baseline model achieves an MAE of $318.70$.

% \begin{figure}
%     \centering
%     \begin{subfigure}{0.5\textwidth}
%         \includegraphics[width=0.4\linewidth]{figs/luts_mlp_learning_curve_58_21.pdf}
%         \caption{CGRA vs. GPGPU.}
%         \label{fig:cgra_gpgpu_tradeoffs} 
%      \end{subfigure}
%      \begin{subfigure}{0.5\textwidth}
%         \includegraphics[width=0.4\linewidth]{figs/luts_xgdb_learning_curve_86_62.pdf}
%         \caption{Performance vs. resource.}
%         \label{fig:perf_vs_rsrc}
%      \end{subfigure}
%      \caption{Tradeoffs between resource, performance, pipelining and parallelization.}
% \end{figure}

%% file: text/results.tex
\section{Evaluation}
\label{results}
We implemented our banking system as a compiler pass in Spatial \cite{spatial}, a fully open-source SDH framework that can be easily modified and target a variety of FPGAs.  
The system uses the Integer Set Library \cite{polyhedral} to perform the polytope emptiness checks.
We evaluated our system on both a Xilinx Virtex 7 and Amazon EC2 F1 instance (Xilinx VU9P) for eight stencil patterns and an additional three patterns from real-world applications described here.

\paragraph{Smith-Waterman} 
SW is a sequence alignment algorithm used in genomics.
It contains a dynamic programming component known as Genome Alignment using Constant-memory Traceback (GACT) \cite{darwin}. 
This essentially creates a sliding-window access pattern where a cell is updated based on the values of its north, west, and north-west neighbor. 
We parallelized this access pattern by 4 to expose wavefront parallelism in the algorithm. 
%Our implementation works on a 512x512 matrix of 8-bit elements.

\paragraph{Sparse Matrix-Vector Multiplication} 
SPMV is a common linear algebra kernel \cite{machsuite}.
Our version uses an edge-list representation to identify dense regions in the matrix.
We parallelized this algorithm over four rows and three columns, so that each row's strided access pattern has a "random" relative offset.
This type of pattern is a good candidate for multidimensional banking because this random offset effectively disappears from projection regrouping (see \ref{projections}).
%Our implementation stores a 64x64 matrix of 32-bit numbers.

\paragraph{Stochastic gradient descent} 
SGD is a training algorithm ubiquitous in machine learning \cite{buckwild}.
Our version uses minibatch, which stores a matrix of input data on-chip and has two modes of access: first in column-major to compute predictions with the model, then in row-major to compute the gradients.
These two modes of access can each be parallelized across rows and columns, and will never be concurrent (i.e. two access groups).  
Our version parallelized both access patterns so that there are 12 accesses in each group.
%Our implementation is on a 64x64 matrix of 64-bit numbers.

\begin{figure}
	\centering
	\begin{subfigure}{0.13\textwidth}
		\centering
		\includegraphics[width=0.95\textwidth]{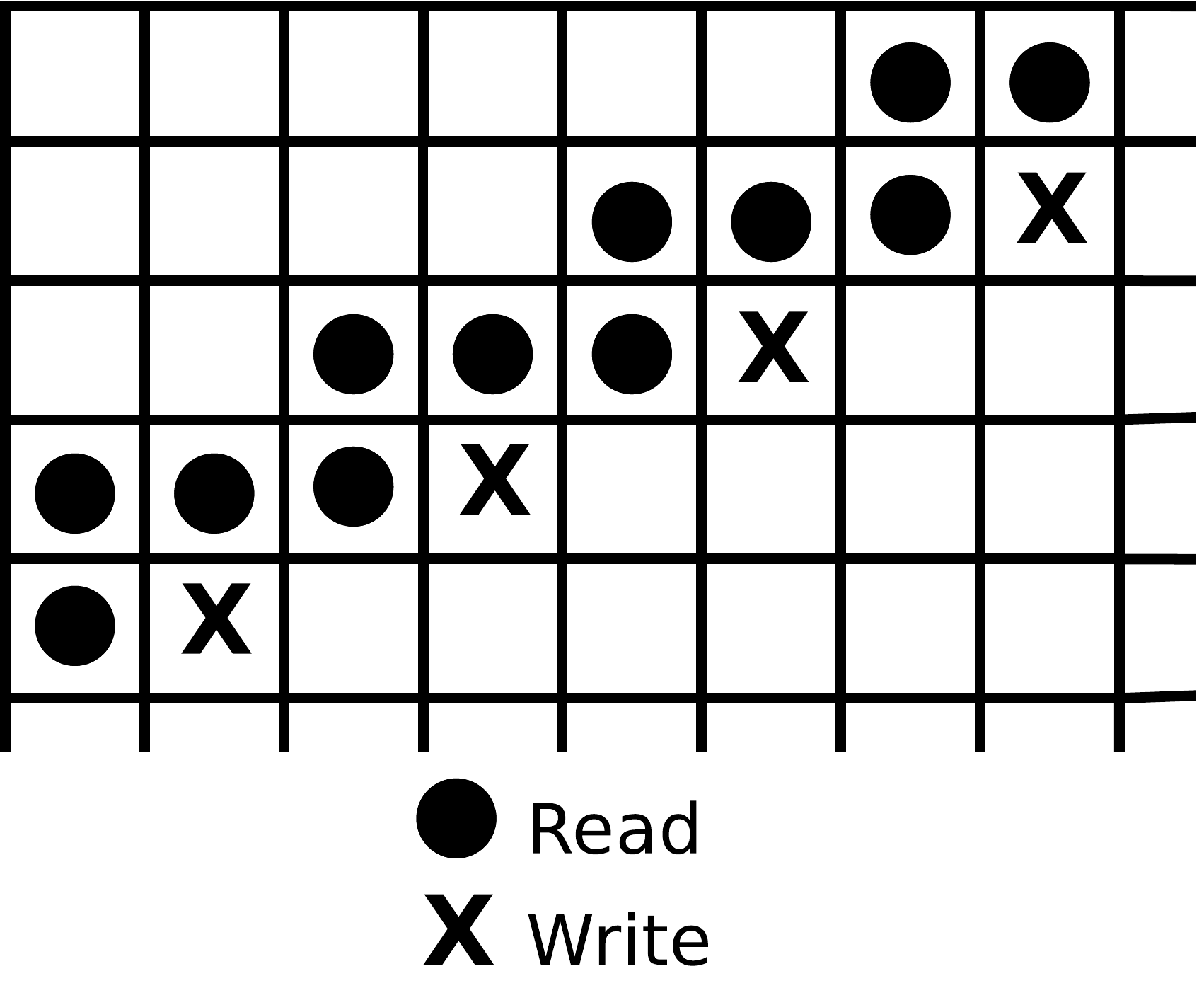}
		\caption{}
		\label{fig:sw}
	\end{subfigure}
	~
	\begin{subfigure}{0.13\textwidth}
		\centering
		\includegraphics[width=0.95\textwidth]{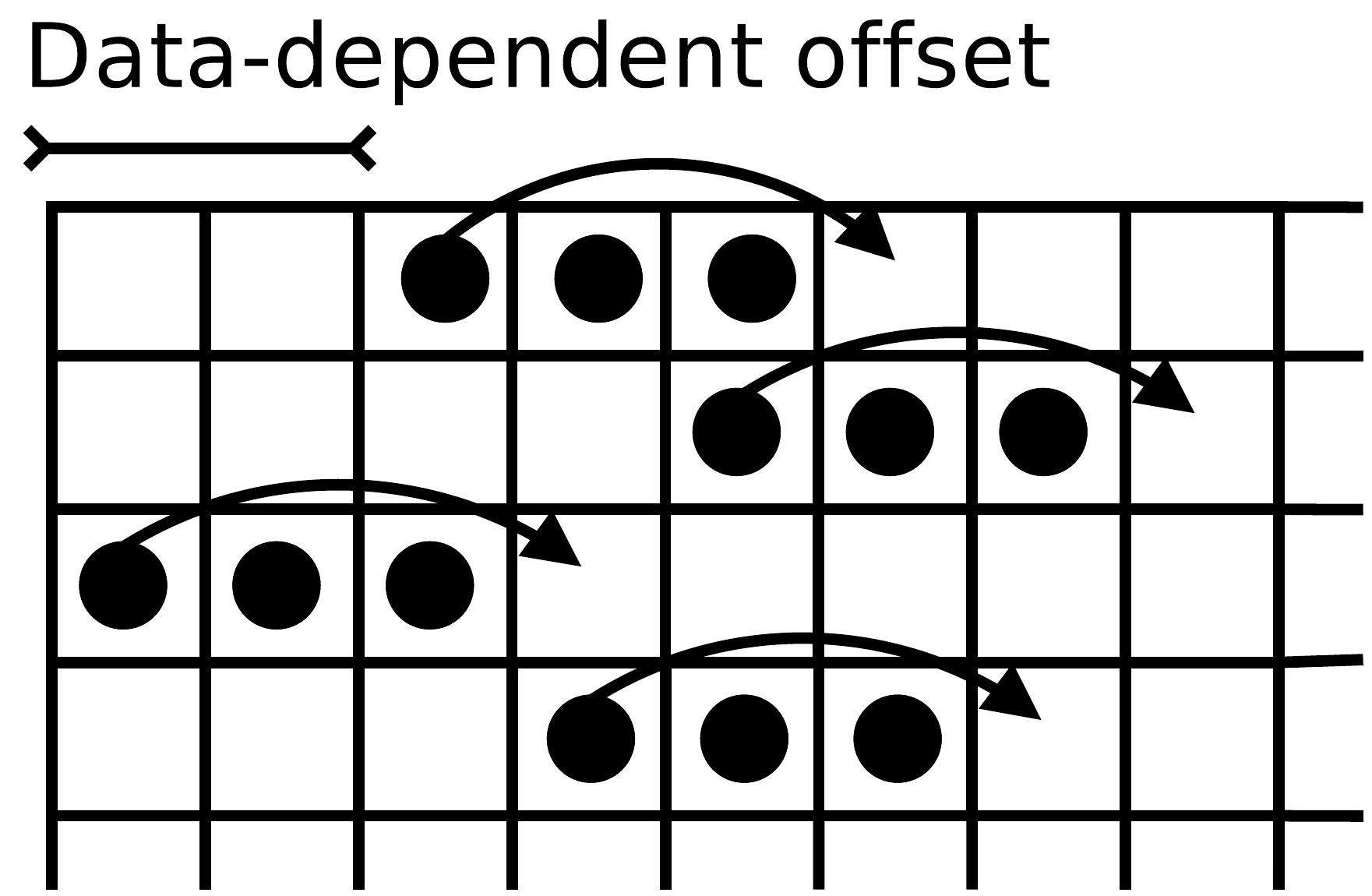}
		\caption{}
		\label{fig:spmv}
	\end{subfigure}
	~
	\begin{subfigure}{0.13\textwidth}
		\centering
		\includegraphics[width=0.95\textwidth]{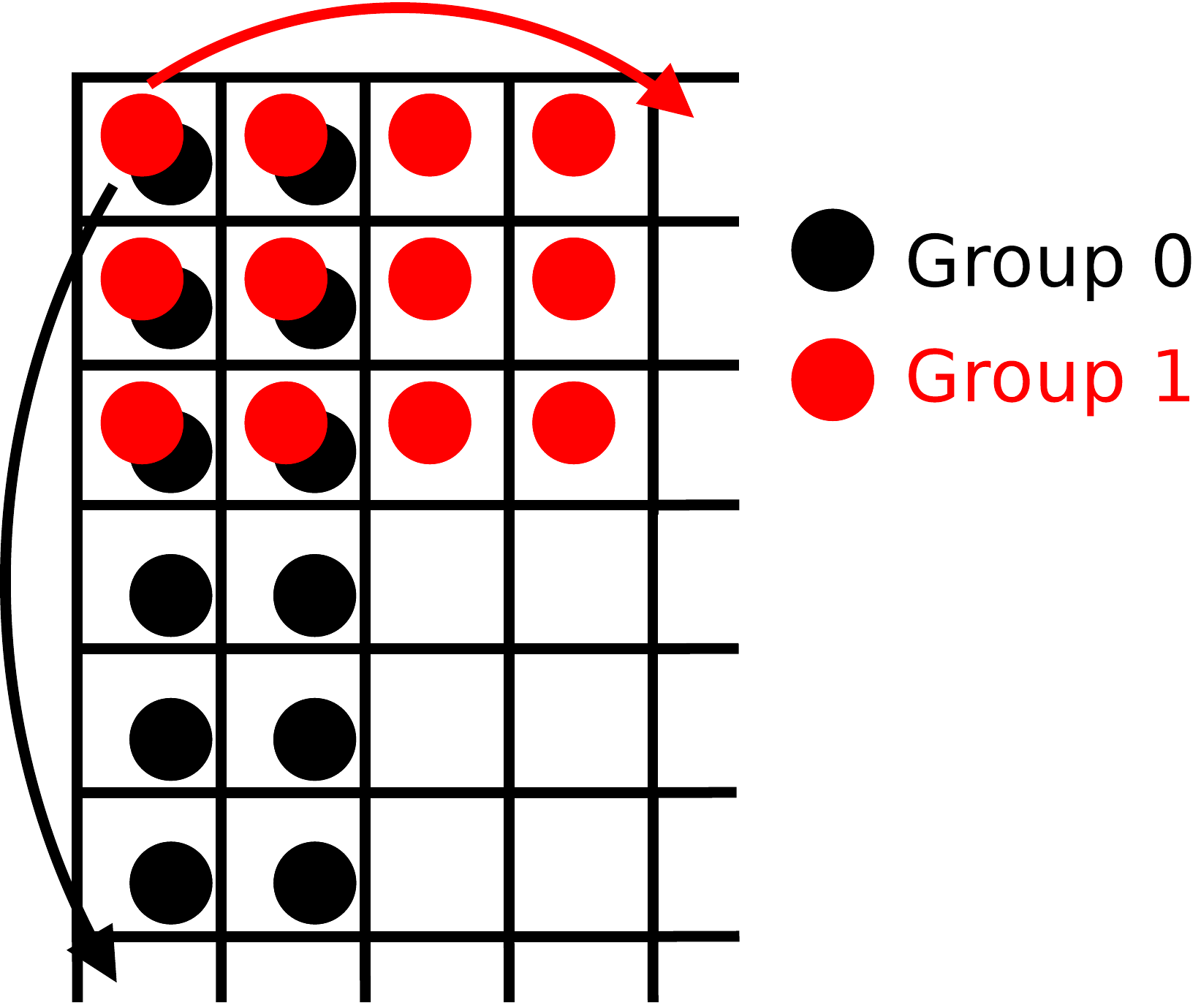}
		\caption{}
		\label{fig:sgd}
	\end{subfigure}
	\caption{Access patterns for sw (a), spmv (b), and sgd (c). Strides are marked in (b) and (c)}
	\label{fig:comparisons}	
	\vskip-.1in
\end{figure}

\subsection{Comparisons on Virtex 7}
\label{simple}
We first compared our system on eight stencil patterns against \cite{cong}, called ``baseline", as well as unmodified Spatial.
Unmodified Spatial uses the first valid scheme it finds.
These results show that solving for numerous solutions and applying resource-saving transformations to each reveals more efficient partitioning schemes in all cases.
Our system always finds parameters that result in DSP-free circuits.;5

%
%
%
%\begin{figure}[h]
%	\centering
%	\includegraphics[width=0.51\textwidth]{figs/cong-eps-converted-to.pdf}
%	\caption{Comparisons for stencil kernels. Lower values are better.}
%	\label{results:cong}
%\end{figure}

\subsection{Comparisons on AWS F1 (VU9P)}
\label{complex}
We also tested our system on a larger FPGA against a state-of-the-art commercial SDH framework called Merlin \cite{merlin} on Amazon's EC2 F1 instances. 
To the best of our knowledge, Merlin does not target the Virtex 7 FPGA so we chose to use the popular F1 backend.
The Merlin and Spatial compiler often land on a partitioning solution that over-utilizes resources.
The key is that our system can view a more diverse set of solutions, take advantage of the constants in the bank resolution arithmetic.
The ML model is what allows our system to choose which one of the many valid solutions will be the best.

For example, the Merlin compiler appears to bank the denoise and bicubic kernels as sobel-like patterns rather than 4-point accesses, hence producing a scheme requiring 9 banks and resource-intensive arithmetic.  
The Spatial compiler detects the ``leaner" solution for these 4-point access patterns by having $\vec{B}\ne1$.
Our system recognizes both of these kinds of solutions, and improves particularly on Spatial's base solution by applying resource-saving transformations on a true dual-ported scheme.

\begin{table}
	\setlength\tabcolsep{4.5pt}
	\begin{tabular}{lllllll}
		\toprule
		\textbf{App} & \textbf{Pattern} & \textbf{System} & \textbf{Slice} & \textbf{BRAM} & \textbf{DSP} \\
		\midrule
		\multirow{3}{0.7cm}{\textbf{denoise}} & \multirow{3}{1cm}{\includegraphics[width=0.035\textwidth]{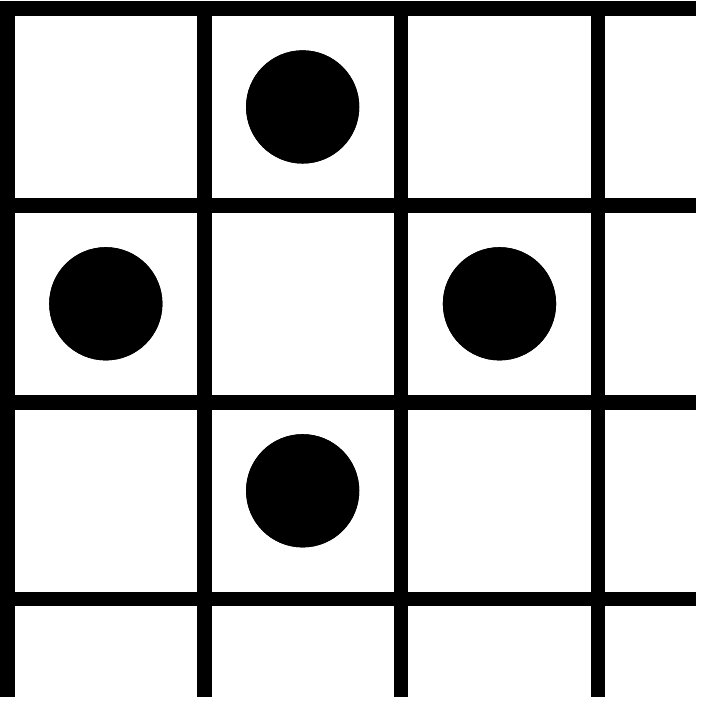}} & Baseline & 303 & 4 & 0 \\
		& & Spatial & 330 & 4  & 0\\
		& & Ours & 213 & 2 & 0  \\
		\midrule
		\multirow{3}{0.7cm}{\textbf{deconv}} & \multirow{3}{1cm}{\includegraphics[width=0.035\textwidth]{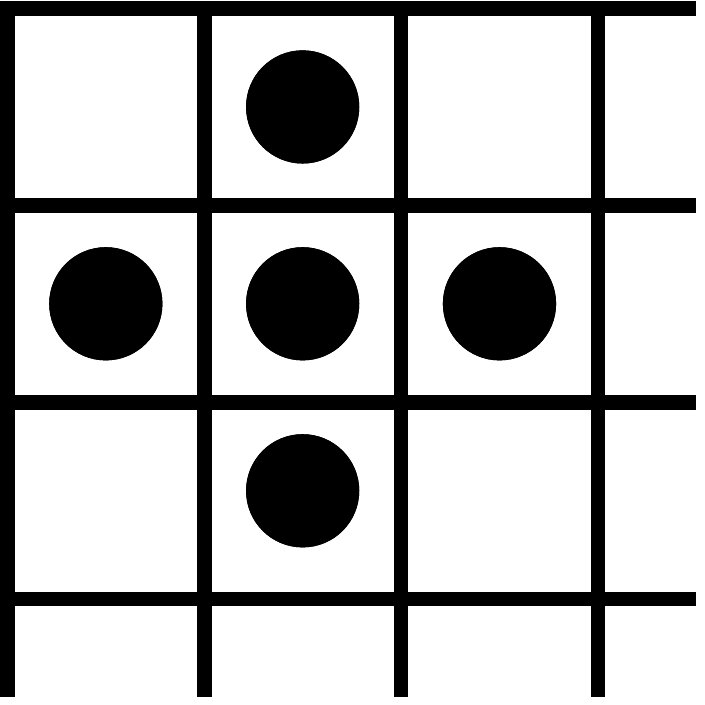}} & Baseline & 597 & 5 & 5 \\
		& & Spatial & 743 & 6 & 4 \\
		& & Ours & 532 & 3& 0 \\
		\midrule
		\multirow{3}{1.7cm}{\textbf{denoise-ur}} & \multirow{3}{1cm}{\includegraphics[width=0.046\textwidth]{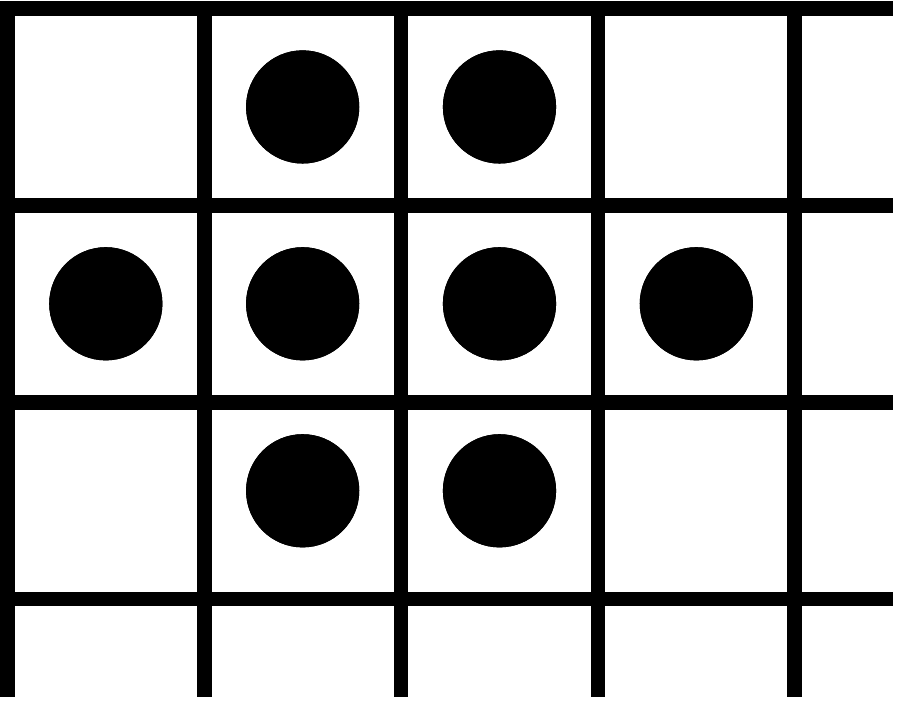}} & Baseline & 795 & 8 & 0\\
		& & Spatial & 1116 & 8 & 5\\
		& & Ours & 659 & 6 & 0\\
		\midrule
		\multirow{3}{0.7cm}{\textbf{bicubic}} & \multirow{3}{1cm}{\includegraphics[width=0.035\textwidth]{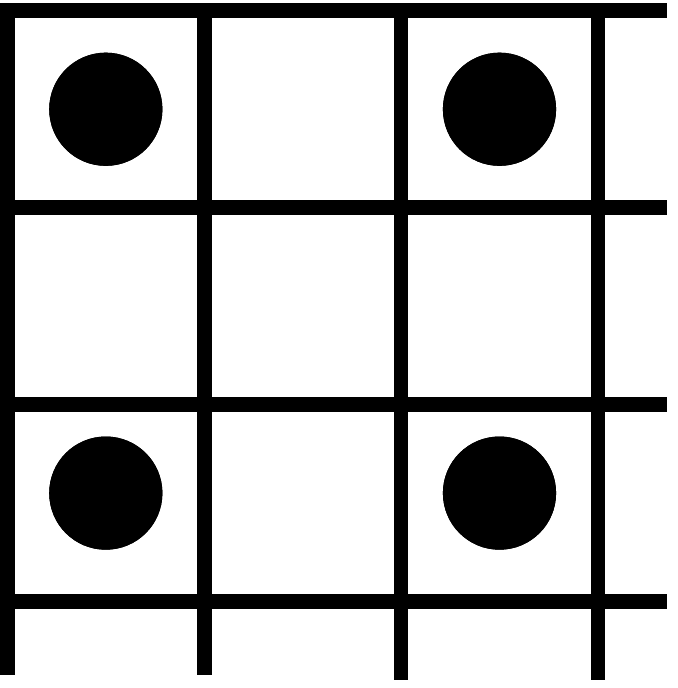}} & Baseline & 238 & 4 & 0 \\
		& & Spatial & 309 & 4 & 0 \\
		& & Ours & 209 & 2 & 0 \\
		\midrule
		\multirow{3}{0.7cm}{\textbf{sobel}} & \multirow{3}{1cm}{\includegraphics[width=0.035\textwidth]{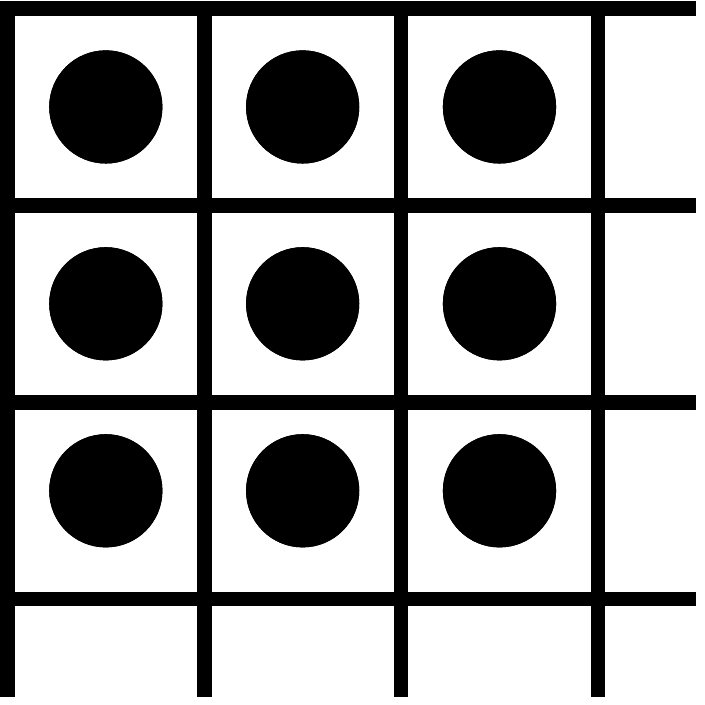}} & Baseline & 1523 & 9 & 9 \\
		& & Spatial & 1801 & 10 & 4 \\
		& & Ours & 1214 & 5 & 0 \\
		\midrule
		\multirow{3}{1.5cm}{\textbf{motion-lv}} & \multirow{3}{1cm}{\includegraphics[width=0.028\textwidth]{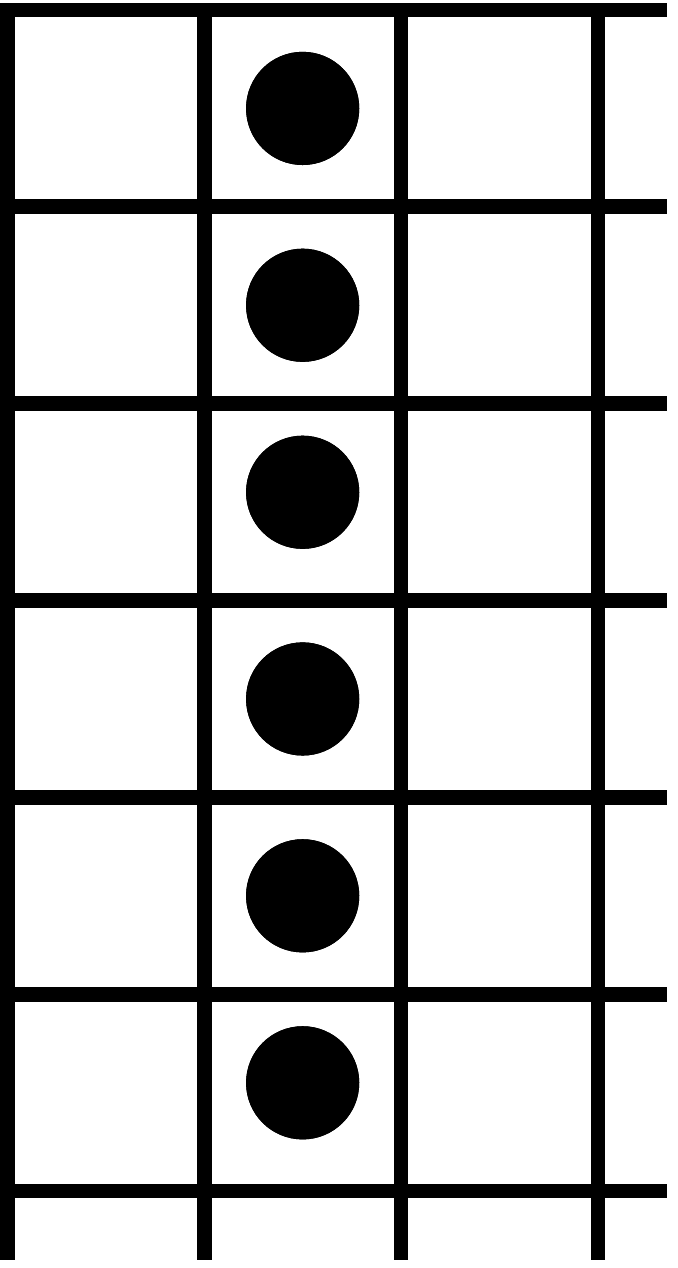}} & Baseline & 425 & 4 & 0 \\
		& & Spatial & 1737 & 6 & 0 \\
		& & Ours & 187 & 3 & 0 \\
		\midrule
		\multirow{3}{1.5cm}{\textbf{motion-lh}} & \multirow{3}{1cm}{\includegraphics[width=0.055\textwidth]{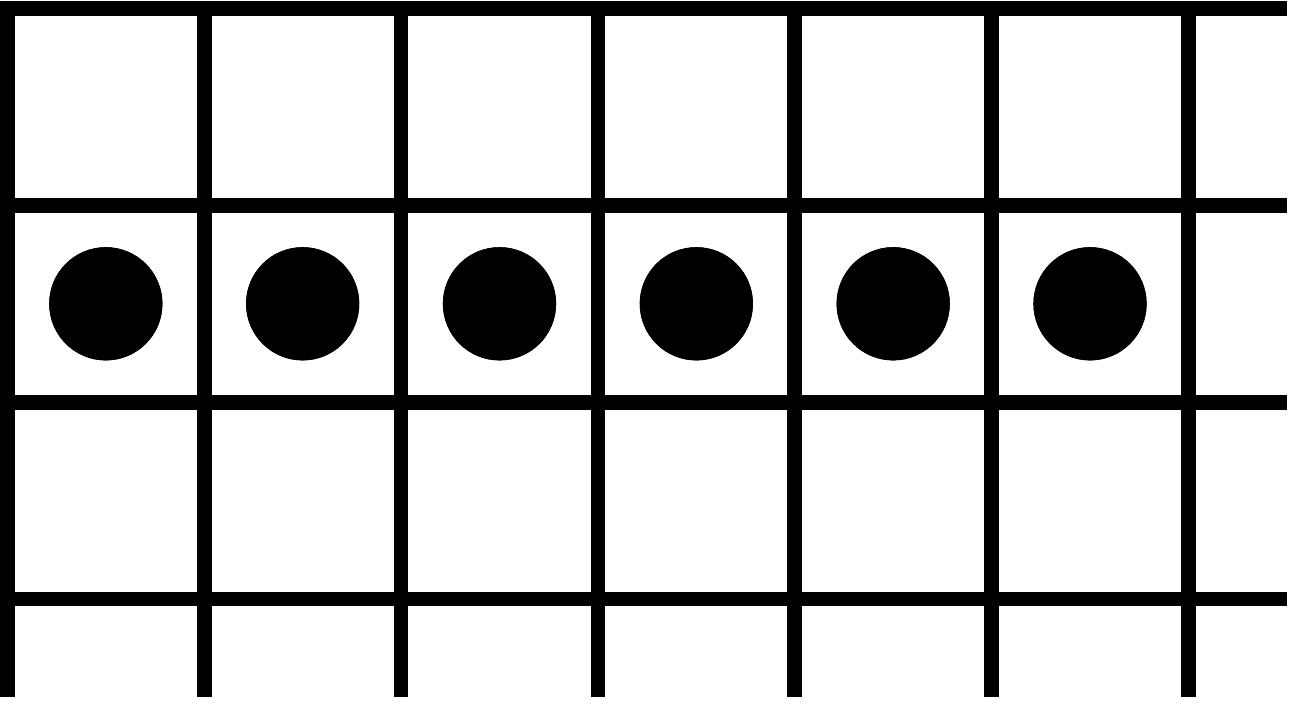}} & Baseline & 334 & 4 & 0 \\
		& & Spatial & 1333 & 6 & 2 \\
		& & Ours & 210 & 3 & 0 \\
		\midrule
		\multirow{3}{1.5cm}{\textbf{motion-c}} & \multirow{3}{1cm}{\includegraphics[width=0.035\textwidth]{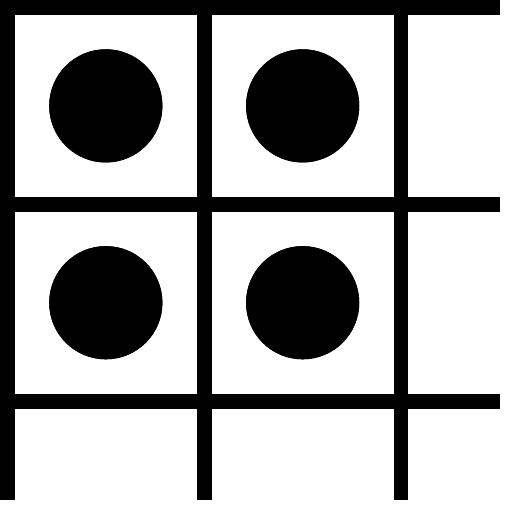}} & Baseline & 155 & 2 & 0 \\
		& & Spatial & 93 & 4 & 0 \\
		& & Ours & 69 & 2 & 0 \\
		\midrule
		\midrule
		\multirow{2}{1.5cm}{Avg. Change} & 
		& Baseline & -29.8\% & -32.4\% & -100\% \\
		& & Spatial & -46.1\% & -46.8\% & -100\% \\		
		\bottomrule
	\end{tabular}
	\caption{Comparisons on Virtex 7}
	\label{fig:virtex}
	\vskip-1.5em
\end{table}

\begin{table}
	\setlength\tabcolsep{2.5pt}
	\begin{tabular}{lllllll}
		\toprule
		\textbf{App} & \textbf{Pattern} & \textbf{System} & \textbf{LUT} & \textbf{FF} & \textbf{BRAM} & \textbf{DSP} \\
		\midrule
		\multirow{3}{0.7cm}{\textbf{denoise}} &  \multirow{3}{1cm}{\includegraphics[width=0.035\textwidth]{figs/denoise-eps-converted-to.pdf}} & Merlin & 4630 & 11523 & 9 & 26  \\
		& & Spatial & 1416 & 1241 & 4  & 0\\
		& & Ours & 184 & 76 & 2 & 0  \\
		\midrule
		\multirow{3}{0.7cm}{\textbf{deconv}} &  \multirow{3}{1cm}{\includegraphics[width=0.035\textwidth]{figs/deconv-eps-converted-to.pdf}} & Merlin & 4795 & 12757 & 9 & 26 \\
		& & Spatial & 2939 & 2537 & 5 & 12 \\
		& & Ours & 2435 & 1119 & 6 & 0 \\
		\midrule
		\multirow{3}{1.7cm}{\textbf{denoise-ur}} &  \multirow{3}{1cm}{\includegraphics[width=0.046\textwidth]{figs/denoiseur-eps-converted-to.pdf}} & Merlin & 3433 & 13342 & 48 & 12\\
		& & Spatial & 2068 & 1240 & 8 & 15 \\
		& & Ours & 638 & 185 & 4 & 0\\
		\midrule
		\multirow{3}{0.7cm}{\textbf{bicubic}} &  \multirow{3}{1cm}{\includegraphics[width=0.035\textwidth]{figs/bicubic-eps-converted-to.pdf}} & Merlin & 4432 & 10788 & 9 & 0 \\
		& & Spatial & 467 & 106 & 4 & 0 \\
		& & Ours & 179 & 65 & 2 & 0 \\
		\midrule
		\multirow{3}{0.7cm}{\textbf{sobel}} &  \multirow{3}{1cm}{\includegraphics[width=0.035\textwidth]{figs/sobel-eps-converted-to.pdf}} & Merlin & 5417 & 17747 & 9 & 24 \\
		& & Spatial & 23157 & 30738 & 9 & 12 \\
		& & Ours & 3482 & 1004 & 9 & 0 \\
		\midrule
		\multirow{3}{1.5cm}{\textbf{motion-lv}} & \multirow{3}{1cm}{\includegraphics[width=0.028\textwidth]{figs/motionlv-eps-converted-to.pdf}} & Merlin & 2894 & 11492 & 6 & 24 \\
		& & Spatial & 19763 & 26706 & 6 & 6 \\
		& & Ours & 1351 & 728 & 3 & 0 \\
		\midrule
		\multirow{3}{1.5cm}{\textbf{motion-lh}} & \multirow{3}{1cm}{\includegraphics[width=0.055\textwidth]{figs/motionlh-eps-converted-to.pdf}} & Merlin & 235 & 289 & 24 & 0 \\
		& & Spatial & 19444 & 26432 & 6 & 0 \\
		& & Ours & 285 & 97 & 3 & 0 \\
		\midrule
		\multirow{3}{1.5cm}{\textbf{motion-c}} &  \multirow{3}{1cm}{\includegraphics[width=0.035\textwidth]{figs/motionc-eps-converted-to.pdf}} & Merlin & 251 & 93 & 16 & 0 \\
		& & Spatial & 241 & 102 & 4 & 0 \\
		& & Ours & 91 & 54 & 2 & 0 \\
		\midrule		
		\multirow{3}{1.5cm}{\textbf{sw}} & \multirow{3}{1cm}{\ Fig \ref{fig:sw}} & Merlin & 6983 & 7625 & 560 & 7 \\
		& & Spatial & 40286 & 48887 & 396 & 21 \\
		& & Ours & 9485 & 3916 & 270 & 0 \\
		\midrule		
		\multirow{3}{1.5cm}{\textbf{spmv}} & \multirow{3}{1cm}{\ Fig \ref{fig:spmv}} & Merlin & 31224 & 11018 & 256 & 0 \\
		& & Spatial & 93972 & 128015 & 20 & 30 \\
		& & Ours & 12269 & 7359 & 6 & 0 \\
		\midrule		
		\multirow{3}{1.5cm}{\textbf{sgd}} & \multirow{3}{1cm}{\ Fig \ref{fig:sgd}} & Merlin & 9466 & 9255 & 504 & 38 \\
		& & Spatial & 21022 & 27725 & 12 & 21 \\
		& & Ours & 4711 & 458 & 12 & 0 \\
		\midrule		
		\midrule
		\multirow{2}{1.5cm}{Avg. Change} & &
		Merlin & -48.1\% & -78.3\% & -71.4\% & -100\% \\
		& & Spatial & -74.0\% & -81.7\% & -37.7\% & -100\% \\		
		\bottomrule	\end{tabular}
	\caption{Comparisons on AWS F1 (VU9P).}
	\label{fig:aws}
	\vskip-1.5em
\end{table}

%% file: text/relatedwork.tex
\section{Related Work}
\label{relatedwork}
The theory of the partitioning problem was pioneered by research in the memory allocation problem in the polyhedral model \cite{memwithpoly} \cite{memreuse} \cite{staticautopar} \cite{autostoragemanagement} \cite{latticealloc} \cite{extendedlatticealloc} \cite{autoarrayopt}.
There is a long history of researchers using this model to build systems that efficiently generate a memory allocation scheme that minimizes the overall memory footprint of an application.
The process is focused on determining the \textit{live-ness} of data in the program and determining a \textit{pseudoprojection} mapping for memory references such that the program can be realized with a smaller memory footprint.

A related line of research involves computing memory partitioning schemes for distributed memory machines (DMMs) \cite{dmm} \cite{commfreedmm} \cite{olddmm}.
In these systems, the compiler builds a partitioning scheme of the program's arrays such that the data is distributed across different parallel processors as efficiently as possible.
The systems use a model that estimates the cost of communicating data between processors to determine the best way to generate the partitioning scheme.

Many SDH frameworks require the programmer to manually partition their arrays, but there has been recent work on developing tools that solve the problem automatically \cite{cong} \cite{offsetcalc} \cite{latticepartition} \cite{powerbanking} \cite{costest} using either hyperplane- or lattice-based techniques.
Either technique can automatically solve the most common partitioning problems, but one may be more efficient for a certain problem than the other.
Furthermore, one technique may have many solutions to the problem and estimating which one is the most cost-effective when synthesized and mapped to hardware can be tricky.
These prior systems generally have first-order rules for determining which solution it finds is the most cost-effective one, but this does not always provide the best answer.
Our system incorporates the advantages of both techniques by looking for both hyperplane and orthogonal lattice solutions.
It also finds alternative parameters for calculating physical addresses, specifically for hyperplane partitioning.
It then uses built-in ML models for estimating the cost of each partitioning scheme when mapped to the underlying FPGA resources.
It also applies mathematical transformations to the bank resolution arithmetic to generate circuits that are more efficient in hardware.
This almost always removes the need for allocating DSPs to the bank resolution logic, which is generally the scarcest resource for algorithms that are accelerated by FPGAs.

%% file: text/conclusion.tex
\section{Conclusion}

As the compiler community continues to develop SDH tools that can express increasingly complicated loop scheduling and parallelization patterns, the problem of automatically partitioning memories quickly and efficiently is becoming increasingly important.
We have introduced a banking analyzer that is scalable with programming models that are equipped to express complicated applications.
Our system provides an appropriate solution to three challenges regarding the problem of memory partitioning: formulating constraints in a flexible SDH framework, quickly finding cheap solutions, and choosing the most cost-efficient one.

First, our system is capable of producing a set of banking constraints regardless of the programmer's choices in loop scheduling, access patterns, and parallelization such that it can guarantee the correctness of its partitioning solution. 
Next, we take advantage of the fact that the bank resolution logic is fully exposed in the IR to implement IR transformations aimed at nodes prevalent in banking resolution arithmetic.
We use heuristics to steer the analyzer towards solutions that can be optimized in hardware.
Finally, we have trained a simple ML model based on decision trees to quickly estimate the resource utilization of laying out various banking schemes in hardware.
The model is accurate enough to select the best scheme.

Overall, all three of these features are necessary to have a practical solution for automatically partitioning memories in the compiler.
Our system is capable of reducing LUT utilization by 86\% and BRAM utilization by 38\% over a naive system for a variety of complicated, real-world benchmarks.
It almost always eliminates all DSP usage from the banking arithmetic for stencil and complex access patterns.
For problems with massive solution spaces, it can cut the time spent searching for solutions in half.